\documentclass[twocolumn]{aastex631}

\usepackage{amsmath}
\usepackage{multirow}
\usepackage{xcolor}
\usepackage{color}

\usepackage{hyperref}
\graphicspath{{../../Figures/}}

\usepackage{graphicx} 
\newcommand\lsim{\mathrel{\rlap{\lower4pt\hbox{\hskip1pt$\sim$}}
\raise1pt\hbox{$<$}}}

\begin{document}
\title{From Binary Wind to Radio Burst: Dispersion-Measure Evolution after Accretion-induced Collapse}
\author[0000-0002-9802-9279]{Smadar Naoz}
\affiliation{Department of Physics and Astronomy, University of California, Los Angeles, CA 90095, USA}
\affiliation{Mani L. Bhaumik Institute for Theoretical Physics, Department of Physics and Astronomy, UCLA, Los Angeles, CA 90095, USA}
\author[0000-0001-9345-0307]{Victoria M.~Kaspi}
\affiliation{Trottier Space Institute, McGill University, 3550 rue University, Montréal, QC H3A 2A7, Canada}
\affiliation{Department of Physics, McGill University, 3600 rue University, Montréal, QC H3A 2T8, Canada}

\begin{abstract}
Secular changes in the dispersion measures (DMs) of repeating fast radio bursts probe the source-local plasma. Such DM time variability has been interpreted with spherical, single-density-profile models. Accretion-induced collapse (AIC) of a white dwarf may instead follow binary interaction and occur inside a finite, anisotropic circumstellar structure. Here, we construct an angle-dependent toy model of aspherical AIC ejecta propagating through a pre-AIC free wind, shocked-wind bubble, swept-up external shell, and polar cavity, exploring a diffuse environment (number density of $1~{\rm cm^{-3}}$, located at $10~$pc) and a compact cloud (number density of $500~{\rm cm^{-3}}$, located at $0.3~$pc), bracketing ionized and initially neutral upstream gas.  The unshocked ejecta initially yield ${\rm DM}\propto t^{-2}$, while crossings of the finite wind-bubble interfaces produce breaks and curvature in ${\rm DM}(t)$.  In initially neutral models, shock ionization can overcome geometric dilution: the total DM reverses from decline to growth after approximately $10$--$20$ yr and later turns over as the finite reservoir is processed. Ionized-upstream models instead generally decline or flatten. 
Without fitting individual sources, our models span both signs and the approximate magnitudes of the secular DM-variation rates measured among repeaters. In this dynamical framework, DM evolution is set by shock propagation through the circumsource density and ionization structure, rather than the contemporaneous burst rate, distinguishing it from photoionization-driven scenarios. Consequently, in our model, an instantaneous \(d{\rm DM}/dt\) is not a unique indicator of source age. We suggest that the observed distribution of secular DM variation slope amplitudes and signs will be important for distinguishing among different models of DM variations in repeating FRBs.
\end{abstract}

\section{Introduction}
Fast Radio Bursts (FRBs) are extremely bright radio transients that last milliseconds and 
exhibit dispersion measures (DMs) exceeding those expected from our Galaxy, indicating that they originate at extragalactic, and usually cosmological distances \citep[e.g.,][]{Lorimer07,Thornton13,Petroff19,CHIME20}. Despite occurring at high rates, estimated at 
around \citep[500~sky$^{-1}$~day$^{-1}$;][]
{CHIME21} in the 400--800-MHz band above 5~Jy~ms,
the underlying nature of FRBs remains a mystery. The observation of an
FRB-like event from the Galactic magnetar SGR 1935+2154 has identified magnetars as a leading candidate for their origin \citep{CHIME20,Bochenek20}. Magnetar-based models can account for many properties of repeaters, including their overall energy budgets \citep[e.g.][]{Abbott+26FRB}, frequent locations in star-forming galaxies \citep[e.g.][]{Tendulkar+17}, as well as accompanying persistent radio sources and large and/or evolving rotation measures  \citep[e.g.,][]{Chatterjee2017,Michilli+2018,sand+2023,Margalit2018,Metzger+19,Margalit+19}. However, FRBs show a wide range of characteristics, including repeating and non-repeating bursts, different polarization properties, and diverse host galaxy types \citep[e.g.,][]{Bannister19,Prochaska19,Heintz20,Bhandari20,Petroff22,Gordon25,Horowicz25,Shah+25}.
There is great diversity even just among repeating sources, with, for example, vast
differences in repeat rates \citep[e.g.][]{Cook+26}.  All
this diversity may reflect multiple progenitor pathways to generate FRBs.

Repeating sources \citep{Spitler2016} are especially interesting to study.
Localization measurements suggest that most FRBs reside in star-forming galaxies \citep{Gordon+23,Sharma+24}. However, notably, some FRBs seem to be associated with older environments.  For example, the repeating FRB~20200120E is associated with a globular cluster in M81 \citep{Bhardwaj+21,Kirsten+22}, and FRB~20240209A lies $\approx40$\,kpc from the center of a quiescent elliptical galaxy \citep{Shah+25,Eftekhari+25}. Old environments like these suggest the possibility of magnetars that form long after star formation has ceased, with time channels with long delays \citep{Kremer+21,Lu+22,Kirsten+22,Shariat+26}.

Accretion-induced collapse (AIC) of a massive white dwarf provides a delayed pathway to forming an FRB-emitting neutron star \citep{Totani13,Kashiyama13,Wang16,Deng21,Sridhar21,Rao25,Shariat+26}. Specifically, \citet{Shariat+26} suggested that AIC triggered via the eccentric Kozai-Lidov (EKL) mechanism \citep{Kozai,Lidov,Naoz16} can naturally produce long delay times in some cases. This triple-induced channel enhances the overall merger rate. In particular, gravitational interactions between the AIC progenitor binary and the tertiary can excite large eccentricities in the inner binary, decreasing its periastron distance and potentially leading to tidal circularization, mass transfer, and/or a stellar merger. Thus, the triple channel often leads to an interacting binary before the AIC, implying that collapse may occur in a non-uniform medium. Notably, in this channel, the pre-collapse mass loss may instead produce a finite and anisotropic circumstellar structure consisting of a freely expanding wind, a wind termination shock, a shocked-wind bubble, a swept-up ambient shell, and the surrounding interstellar medium \citep[e.g.,][]{Castor+75,ChevalierSoker89,Klein+94,Hillel+20,Hillel+22,Hillel+26}.

Because the Milky Way, intergalactic, and large-scale host-galaxy DM 
contributions should be 
small on year-long timescales -- 
Galactic pulsar timing shows changes of at most $\sim$0.15~pc cm$^{-3}$~yr$^{-1}$ (e.g., Kaspi et al. 1994; Petroff et al. 2013), which we take as representative of the host ISM, while the halo and IGM contributions vary even less \citep[e.g][]{Yang+17} --
secular changes in the DM 
of a repeating FRB provide a sensitive probe of plasma local to the source \citep[e.g.,][]{Piro16,Piro+18,Yang+17,Zhao+21}. Such DM evolution has now been measured for several repeaters \citep[e.g.,][]{Cook+26,Abbott+26FRB,Jia+26,Pandhi2026}. Notable examples include FRB~20121102A, whose DM increased for several years before reversing into a decline \citep{Hessels2019,WangP2025,Snelders+25,Waxman+26}; FRB~20190520B, which has an unusually large DM and one of the most rapid secular declines yet measured \citep{Niu2022,Niu+26}; FRB~20220529A, which exhibits a steady long-term decline superposed with a short-lived DM excursion \citep{Pandhi2026}; and
FRB~20220912A shows steadily increasing DM 
\citep{Abbott+26FRB}.
\citet{Cui2026} reports tentative population-level evidence for declining trends, but this still needs verification.

Secular DM variability is commonly attributed to dynamical evolution of the source-local plasma: expanding ejecta become diluted, forward and reverse shocks ionize and redistribute material, and the forward shock sweeps up the surrounding medium. Analytic studies have generally treated these processes in spherical symmetry, considering ejecta expanding into either a uniform ambient medium or an unbroken $r^{-2}$ wind \citep[e.g.,][]{Piro+18,Kundu2020,Zhao+21}. More recent numerical studies have relaxed some of these assumptions. For example, \citet{Zhang+26} followed the evolution in one dimension through circumstellar media shaped by progenitor mass loss, whereas \citet{Fang2026} modeled an anisotropic neutron-star wind inside otherwise unstructured supernova ejecta in two dimensions and found moderate viewing-angle variations in the DM. The dynamical interpretation is not unique, however. \citet{Metzger26} suggests that changes in the photoionization state of the surrounding material can drive the observed DM variability, while \citet{Waxman+26}, combining the DM evolution and persistent radio source of FRB~20121102A, find the confining plasma difficult to reconcile with typical supernova ejecta and suggest a white-dwarf merger as a possible origin. 
{{
 \citet{Ocker+26Kick} argued that source motion along the line of sight changes the path length through local ionized gas, and transverse motion samples fluctuations in the electron column, both of which can result in secular DM variations. For reasonable source space velocities, large ambient electron densities are required in that picture.

}}

Here, we construct an angle-dependent toy model for the local DM of an FRB source formed through AIC inside a binary-sculpted circumstellar medium. Unlike calculations with a single power-law external density, we follow aspherical ejecta through a finite free-wind region, the pre-AIC wind termination shock and shocked-wind bubble, the swept-up ambient shell, and the surrounding ISM, while allowing an evacuated polar channel. We track the electron column by both its present spatial location and its material origin. Our goal is to identify how ejecta properties, wind anisotropy, ambient density, ionization, and viewing angle determine \({\rm DM}_{\rm loc}\) and \(d{\rm DM}_{\rm loc}/dt\). Interface crossings robustly produce viewing-angle-dependent changes in component slopes, whereas a rise or turnover in the total DM depends on the ionization contrast and the dominant component. Fitting a spherical model to a single source can then misestimate its age and environment.

The paper is organized as follows: we describe the multiple components in Section \ref{sec:ToyModel}, calculate the different DM components in Section \ref{sec:zones}, and the DM time variability in Section \ref{sec:DMvar}. We finally offer our discussion in Section \ref{sec:diss}.

\begin{figure*}[t]
    \centering
    \includegraphics[width=0.95\linewidth]{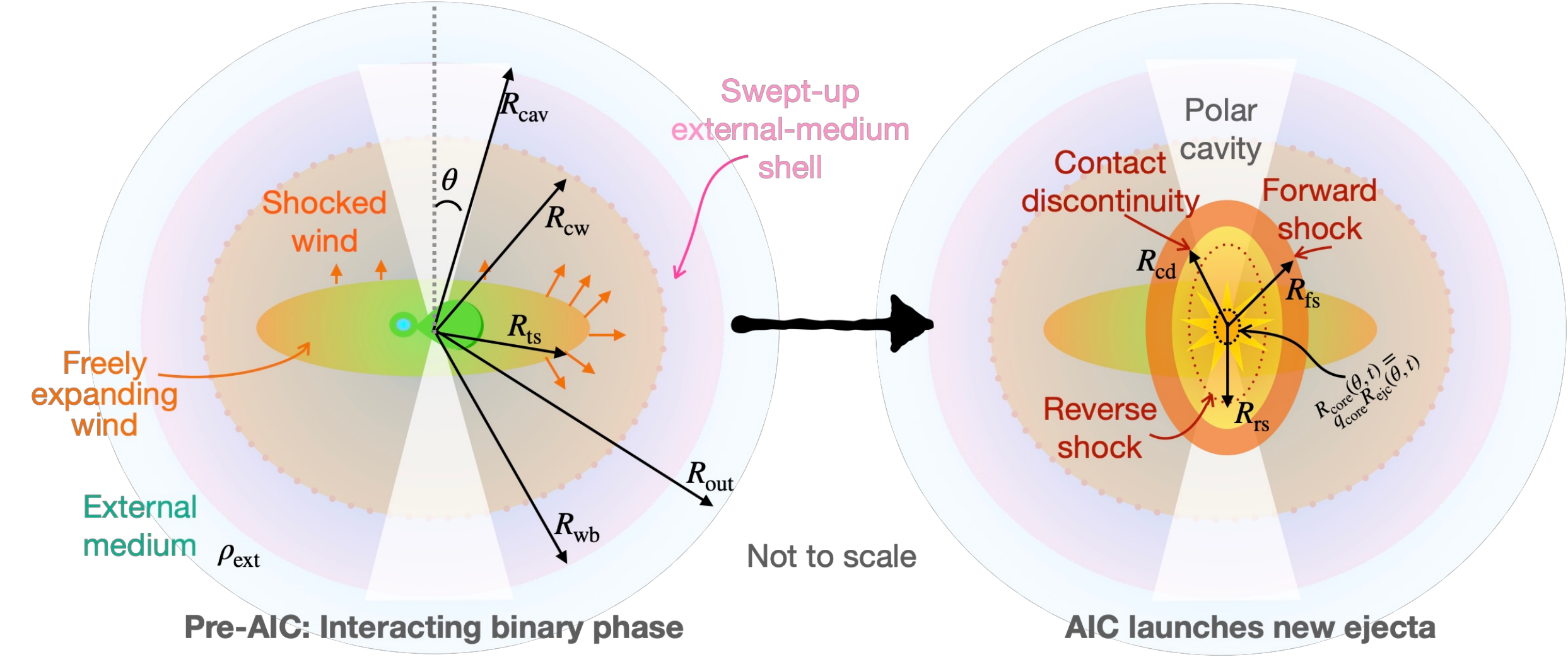}
    \caption{Illustration of the pre-AIC interacting binary system (left) and early AIC ejecta––circumstellar interaction (right). The cartoon is not to scale. The black dashed contour marks the Lagrangian core--envelope break,
\(q=q_{\rm core}\). While this material remains unshocked, its
physical radius along each direction is
\(R_{\rm core}(\theta,t)=q_{\rm core}R_{\rm ejc}(\theta,t)\).
It is a material boundary rather than a shock.  }
    \label{fig:Cartoon}
\end{figure*}

\section{The Toy Model}\label{sec:ToyModel}

\subsection{Overview of the model}
In the pre-AIC era, the binary continually injects wind, which inflates a pressurized bubble. We define $t_w$ as the duration/age of the pre-AIC wind bubble. We adopt a pre-collapse duration of the later phase of the fast wind, $t_w=5000$~yr, which is at the lower limit of the expected range\footnote{We also tested $t_w=10^5$~yr, for slow $100$~km~sec$^{-1}$ and fast $1000$~km~sec$^{-1}$ winds, both of which yield qualitatively similar results, omitted here to avoid clutter.  In these tests we required the dense cloud outer edge to lie outside the pre-AIC wind bubble.   } for pre-AIC mass-transfer phases of few$\times 10^{3-5}$ yr \citep[e.g.,][]{Brooks+17, Wang18,Yu+19}. 
This wind inflates a pressurized bubble in the ambient medium, bounded on the outside by a shock running into the external medium, and on the inside by a wind termination shock, $R_{\rm ts}$ \citep[e.g.,][]{Castor+75,Weaver+77}. 
Time $t$ represents the time after the AIC ejecta is launched and expands homologously into the pre-existing structure. Thus, before the AIC forward shock reaches the termination shock, the sequence of interfaces, in the non-polar direction, is (see illustration in Figure \ref{fig:Cartoon}):
\begin{equation}
  0 < R_{\rm rs} < R_{\rm cd} < R_{\rm fs} < R_{\rm ts} < R_{\rm cw} < R_{\rm wb} < R_{\rm out},
  \label{eq:ordering}
\end{equation}
where $R_{\rm rs}$, $R_{\rm cd}$ and $R_{\rm fs}$ are the reverse shock, contact discontinuity and forward shock of the AIC ejecta--circumstellar interaction, $R_{\rm ts}$ is the pre-AIC wind termination shock, $R_{\rm cw}$ is the contact discontinuity between shocked wind and swept-up ISM, $R_{\rm wb}$ is the outer shock of the wind bubble, and $R_{\rm out}$ is the outer edge of the modeled local environment. In the polar direction, we also introduce the cavity radius, $R_{\rm cav}$ (see \S \ref{sec:polar_cavity} and Appendix \ref{app:early_wind}). All radii depend on $\theta$, measured from the symmetry axis (the orbital angular momentum of the binary). 
 
We place the FRB engine at the origin\footnote{See \S \ref{sec:diss} for discussion on this point. } and calculate the local dispersion measure, ${\rm DM}_{\rm loc}(t,i)$, as a function of time since collapse, $t$, and the viewing angle, $i$, measured from the symmetry axis. Because the source is at the origin, the sightline to an observer at inclination $i$ is the radial ray $\theta=i$, so the polar coordinate $\theta$ and the viewing angle $i$ coincide in all expressions below.

The observed DM of a burst at redshift $z$ is
\begin{equation}
  {\rm DM}_{\rm obs} = {\rm DM}_{\rm MW} + {\rm DM}_{\rm halo} + {\rm DM}_{\rm IGM}(z)
  + \frac{{\rm DM}_{\rm host} + {\rm DM}_{\rm loc}}{1+z},
  \label{eq:dmobs}
\end{equation}
where ${\rm DM}_{\rm host}$ is the contribution of the host galaxy outside the local environment \citep[e.g.,][]{Deng+14,Macquart+20,Zhao+21}.
Neither DM$_{\rm halo}$ nor DM$_{\rm IGM}$ are expected to vary on human timescales \citep[e.g.][]{Yang+17}, and from pulsar observations in our Galaxy, we expect changes of at most
$\sim 0.15$~pc~cm$^{-3}$~yr$^{-1}$ from DM$_{\rm MW}$ or DM$_{\rm host}$ \citep[e.g.][]{petroff13}, leaving DM$_{\rm loc}$ as the lone source of variation comparable to observed trends, so
we focus on that contribution. 
The free-electron density is
\begin{equation}
  n_e(r,\theta,t)=\frac{x_e\,Y_e}{m_p}\,\rho(r,\theta,t),
  \label{eq:ne}
\end{equation}
where $Y_e$ is the number of electrons per baryon when fully ionized and $x_e$ is the ionized fraction.  Realistically, the ionization fraction evolves with time \citep[e.g.,][]{Metzger26}. However, throughout the manuscript we hold parameters constant and bracket the ionization level with two scenarios.   In the first, we assume that both upstream and downstream regions are ionized; in the second, we adopt a partly ionized model in which the upstream is neutral, corresponding to the unshocked pre-AIC swept-up material and the unshocked ambient external material.  For the unshocked ejecta, we assume $x_{e,{\rm ejc}}=0.1$ and $Y_{e,{\rm ejc}}=0.5$, where $x_e=0.1$ is adopted to represent a low ionization state value within the range commonly considered in FRB ejecta models
\citep{Kundu2020,Zhao+21}. Similarly, $Y_e=0.5$ corresponds to a material dominated by nuclei with $Z/A\simeq1/2$
and is obtained in neutrino-irradiated AIC disk outflows
\citep{Metzger+09}. However, AIC calculations predict a broad,
and potentially angle-dependent, distribution of $Y_e$, including
more neutron-rich ejecta \citep{Dessart+06,Dessart+07}.  

\subsection{The pre-AIC Circumstellar Environment}\label{sec:csm}

We adopt the standard zonal structure of a wind-blown bubble, consisting of (from inside out) a freely expanding wind, a wind-termination shock, a shocked-wind interior, and a swept-up shell \citep[e.g.,][]{Castor+75Wind,Weaver+77}.
We model the wind as: 
 \begin{equation}\label{eq:rhow}
        \rho_{\rm w} (r,\theta) = K(\theta) r^{-2} \quad R_{\rm w,in} <r<R_{\rm ts}(\theta) \ ,
    \end{equation}
    where we adopt for the wind ambient-density profile, $r^{-s}$, a power-law index $s=2$ throughout the paper. The $r^{-2}$ profile assumes a steady, radial wind that has reached its terminal velocity. Note that we impose an inner cutoff \(R_{\rm w,in}=1\,{\rm au}\) on the free-wind profile. This radius 
    is on a scale comparable to the binary separation; mass transfer and wind--companion interaction produce a time-dependent, non-radial flow, so that a steady \(K(\theta)r^{-2}\) profile centered on the binary is not expected to apply. The adopted value is characteristic of wide, symbiotic or recurrent-nova-like systems\footnote{ We emphasize that \(R_{\rm w,in}=1\,{\rm au}\) is a fiducial choice rather than a prediction for all AIC progenitors, some of which may be substantially
more compact \citep[e.g.,][]{Brooks+17}.}. For example, RS~Oph has an orbital period of \(453.6\)~d and component masses that imply a separation of approximately \(1.5\)~au \citep{Walder+08,Brandi+09}. 
    $K(\theta)$ is then:
    \begin{equation}\label{eq:Ktheta}
        K(\theta) = \frac{d\dot{M}/d\Omega(\theta)}{v_w(\theta)} \ ,
    \end{equation}
    where $d\dot{M}/d\Omega(\theta)$ is the mass loss rate per solid angle and $v_w(\theta)$ is the speed of the wind. 
   We adopt a terminal wind speed \(v_w=1000\ \mathrm{km\,s^{-1}}\), representative of the lower end of fast outflows from accreting white dwarfs and nova systems, for which characteristic velocities span approximately \(10^3\) to a few\(\times10^3\ \mathrm{km\,s^{-1}}\) \citep[e.g.,][]{Prinja+95,Hachisu+96,Chomiuk2014}. This parameter describes the fast WD/accretion-driven component rather than the slower donor wind or binary-driven equatorial outflow\footnote{We tested the sensitivity of the results to different values of $v_w$ and find that the qualitative results are consistent throughout the mentioned range, and even for winds as slow as $\sim 100$~km~s$^{-1}$, associated with $t_w=5000$~yr. We discuss these tests in \S \ref{sec:diss}.  }.  Within the imposed polar cavity, we set the wind density and all wind-origin DM contributions to zero, and treat the cavity as not evolving as a function of time until the AIC ejecta reach it. 

 The mass loss from binary interaction can concentrate the slower, denser outflow toward the orbital plane, while faster WD- or disk-driven winds may be more nearly isotropic or polar \citep[e.g.,][]{Walder+08,Shen+22,Hillel+22}. 
We describe the angular distribution of the wind with the lowest-order axisymmetric (Legendre $P_2$) term,
\begin{equation}
  \frac{d\dot M}{d\Omega}(\theta)=\frac{\dot M}{4\pi}
  \left[1+\frac{\epsilon}{2}\left(3\cos^2\theta-1\right)\right].
  \label{eq:mdot}
\end{equation}
Positivity requires $-1<\epsilon<2$; $\epsilon<0$ gives an equatorially enhanced (oblate) outflow.  
Our fiducial value for $\epsilon$ is motivated by 
\citet{Walder+08} simulations of the recurrent nova RS Oph. They found an orbital-plane density enhancement of a factor of $\sim 2-3$, during quiescence, on scales beyond the binary interaction radius, with substantially larger, non-axisymmetric contrasts closer to the stars. 
Equation~(\ref{eq:mdot}) specifies a smooth reference wind before imposing the polar cavities. At a common radius, the nominal equatorial and polar densities of this reference profile satisfy
\begin{equation}
   \frac{\rho_{\rm eq}}{\rho_{\rm pol}}
=\frac{K(\pi/2)}{K(0)}
=\frac{1-\epsilon/2}{1+\epsilon} \ . 
\end{equation}
Thus, a nominal equator-to-pole density contrast of two implies $\epsilon=-0.4$. For our angle-independent wind speed and uniform bubble pressure, pressure balance gives $R_{\rm ts}\propto[(d\dot{M}/d\Omega)v_w/P_{\rm b}]^{1/2}$, and hence a nominal termination-radius ratio $R_{\rm ts,eq}/R_{\rm ts,pol}=\sqrt{2}$. Here, $\rho_{\rm pol}$ and $R_{\rm ts,pol}$ are formal reference values obtained by extending the smooth wind prescription to the axis.
The adopted density profile instead imposes an evacuated polar cavity, with $\rho_{\rm cav}\simeq0$ and no wind termination shock inside the cone (Section~\ref{sec:polar_cavity} and Equation ~\ref{eq:pre_aic_density_profile}). Accordingly, $\epsilon$ controls the smooth wind anisotropy outside the cavities, while $\theta_{\rm cav}$ independently specifies their opening angle. We tested different nominal contrast values and found qualitatively similar results.


The shocked-wind region, $R_{\rm ts}<r<R_{\rm cw}$, contains the wind emitted over
$t_w$ minus the mass still in free expansion. Assuming mass conservation and uniform density within each angular sector, we can write: 
\begin{equation}
  \frac{dM_{\rm sw}}{d\Omega}=\frac{d\dot M}{d\Omega}\,t_w-K\left(R_{\rm ts}-R_{\rm w,in}\right) \ ,
\end{equation}
where $K=K(\theta)$, defined in Eq.~(\ref{eq:Ktheta}), and
\begin{equation}\label{eq:rhosw}
     \rho_{\rm sw}(\theta)=\frac{3\,dM_{\rm sw}/d\Omega}{R_{\rm cw}^3-R_{\rm ts}^3}
\end{equation}
The subscript ``sw'' means pre-AIC shocked-wind origin.
The ISM, or external medium material, originally inside $R_{\rm wb}$ is swept into the shell $R_{\rm cw}<r<R_{\rm wb}$,
\begin{equation}
  \frac{dM_{\rm shell}}{d\Omega}=\frac{\rho_{\rm ext}R_{\rm wb}^3}{3} \ ,
  \label{eq:dMrhosh}
\end{equation}
{{As a geometric approximation, we take the outer wind-bubble to be spherical}} 
(i.e., $R_{\rm wb}$ and $R_{\rm cw}$ are independent of $\theta$), consistent with a uniform interior pressure; only the free-wind region and $R_{\rm ts}$ carry the binary anisotropy. In other words, we adopt that ${dM_{\rm shell}}/{d\Omega}$ is constant across the non-polar region.  

\subsection{Polar Cavity and External Medium}
\label{sec:polar_cavity}

In addition to the equatorially enhanced wind, we assume that a collimated, jet-like component of the pre-AIC binary outflow evacuates two polar channels (see illustration in Figure \ref{fig:Cartoon}). 
Within these directions, the residual
density is assumed to be dynamically negligible out to the cavity wall, $R_{\rm cav}$. The polar ejecta therefore encounter the external medium directly, without first crossing a freely expanding wind, a wind termination shock, a shocked-wind bubble, or a pre-AIC swept-up shell. While, in general, $R_{\rm cav}$ can be dependent on $\theta$, to simplify the toy model, we approximate that  $R_{\rm cav}= R_{\rm wb}$, where $R_{\rm wb}$ represents the region where the external material has been swept up by the outer forward shock. Having $R_{\rm cav}< R_{\rm wb}$ can lead to further DM variability at the polar viewing angle. We omit these tests here to avoid clutter, especially because we verified that these effects remain relatively small. 

Combining the polar and non-polar prescriptions, the adopted pre-AIC density profile is
\begin{equation}
\rho_{\rm pre}(r,\theta)=
\begin{cases}
\rho_{\rm cav}\simeq 0,
&
\theta\leq\theta_{\rm cav},
\quad
0\leq r<R_{\rm cav},
\\[3pt]
\rho_{\rm ext},
&
\theta\leq\theta_{\rm cav},
\quad
R_{\rm cav}\leq r<R_{\rm out},
\\[3pt]
\rho_{\rm w}(r,\theta),
&
\theta>\theta_{\rm cav},
\quad
R_{\rm w,in}\leq r<R_{\rm ts}(\theta),
\\[3pt]
\rho_{\rm sw}(\theta),
&
\theta>\theta_{\rm cav},
\quad
R_{\rm ts}(\theta)\leq r<R_{\rm cw},
\\[3pt]
\rho_{\rm shell},
&
\theta>\theta_{\rm cav},
\quad
R_{\rm cw}\leq r<R_{\rm wb},
\\[3pt]
\rho_{\rm ext},
&
\theta>\theta_{\rm cav},
\quad
R_{\rm wb}\leq r<R_{\rm out} \ ,
\end{cases}
\label{eq:pre_aic_density_profile}
\end{equation}
where $\rho_{\rm sw}$ is the averaged density of wind-origin material that crossed the pre-AIC termination shock, and $\rho_{\rm shell}$ is the density of the material originally inside $R_{\rm wb},$ that is swept into the shell.
Here $\rho_{\rm ext}=m_p n_{\rm ext}$ is the density of the external medium, which may represent either a diffuse ISM or a dense cloud. The free-expanding wind $\rho_w(\theta)=K(\theta)r^{-2}$, is defined in Equation (\ref{eq:rhow}). The quantities $\rho_{\rm sw}(\theta)$ and $\rho_{\rm shell}$ are, respectively, the adopted mean densities of the pre-AIC shocked-wind region and swept-up external-medium shell. See Equation~(\ref{eq:rhosw}) for $\rho_{\rm sw}(\theta)$, while the shell of the swept-up external-medium in the shell $R_{\rm cw}<r<R_{\rm wb}$, yields: 
\begin{equation}\label{eq:rhosh}
  \rho_{\rm shell}=\frac{\rho_{\rm ext}}{1-\xi^3} 
\end{equation}
where $\xi = R_{\rm cw}/R_{\rm wb}=0.86$, taken from the \citet{Castor+75Wind}  the \citet{Weaver+77} wind-bubble solution.  

We adopt $\theta_{\rm cav}=20^\circ$ as a representative narrow polar channel. Jet-driven, strongly nonspherical outflows are plausible in interacting binaries \citep[e.g.,][]{Hillel+20,Hillel+22}, but their
opening angles are system-dependent; thus, $\theta_{\rm cav}$ should be regarded as a phenomenological parameter rather than a universal prediction. For our fiducial wind anisotropy, $\epsilon=-0.4$, the two polar cones contain approximately $3.8\%$ of the nominal mass-loss rate implied by Equation~(\ref{eq:mdot}). We neglect this small fraction rather than renormalizing the wind outside the cavities.

In polar directions, the AIC ejecta consequently expand freely until their outer edge reaches the cavity wall at
\begin{equation}
    t_{\rm hit}(\theta)
    =
    \frac{R_{\rm cav}}{v_{\rm ejc}(\theta)} \ ,
    \label{eq:polar_hit_time}
\end{equation}
where \(v_{\rm ejc}(\theta)\) is the direction-dependent outer ejecta velocity defined below. 
For $t<t_{\rm hit}$, no forward shock, contact discontinuity, or reverse shock is present along that direction. At $t=t_{\rm hit}$, the
ejecta impact the external medium and launch a forward shock.

As a proof of concept, we consider two contrasting external environments, without attempting to fit an individual repeater. Our diffuse, extended model adopts \(n_{\rm ext}=1~{\rm cm^{-3}}\) and \(R_{\rm out}=10~{\rm pc}\), placing the outer boundary of the modeled source environment well beyond the pre-AIC wind bubble. Our compact, dense model is motivated by \citet{Waxman+26}, who infer that the persistent radio source associated with FRB~20121102A is confined by cold plasma on scales of \(R\sim10^{17}\)--\(10^{18}~{\rm cm}\), or approximately \(0.03\)--\(0.3~{\rm pc}\). We therefore adopt {for this case} \(n_{\rm ext}=500~{\rm cm^{-3}}\) and \(R_{\rm out}=0.3~{\rm pc}\). We note that \(R_{\rm out}=0.3~{\rm pc}\) is set outside $R_{\rm wb}$ for the fiducial values chosen here. We also tested other values of $R_{\rm out}$ associated with different $t_w$, specifically $t_w=10^4$~yr and $t_w=10^5$~yr, which give $R_{\rm out}\sim 0.4$~pc and $1.6$~pc, respectively, and obtained qualitatively similar results, omitted here to avoid clutter.  Before accounting for the excavation and redistribution of material by the pre-AIC wind bubble, these parameters correspond to a characteristic column \(n_{\rm ext}R_{\rm out}=150~{\rm pc\,cm^{-3}}\), comparable in order of magnitude to the source-local column inferred for FRB~20121102A. This model is intended as an illustrative compact environment rather than a direct implementation or fit of the \citet{Waxman+26} model. See Table \ref{tab:modelparamters} for the model parameters. 




\begin{table*}[t]
    \centering
    \begin{tabular}{c|ccccc|l}
      Name &  ${\rm M}_{\rm ejc}$ & $E_k$ & Upstream & $R_{\rm out}$ & $n_{\rm ext}$ & Lines in \\
         &        M$_\odot$ & erg         &  ionization & pc & cm$^{-3}$ & Figs.~\ref{fig:Dynamics}-\ref{fig:SlopesHist}
         \\  \hline
        ISM & $0.1$ &  $10^{50}$ &  Ionized & $10$ & $1$ & Solid line\\ 
        ISM & $0.01$ &  $10^{49}$ &  Ionized & $10$ &$1$ & Dashed line \\
        ISM & $0.1$ &  $10^{50}$ &  Neutral & $10$ &$1$ & Solid line\\ 
        ISM & $0.01$ &  $10^{49}$ &  Neutral & $10$ & $1$ & Dashed line \\
        Dense cloud & $0.1$ &  $10^{50}$ & Ionized & $0.3$& $500$ & Solid line\\
         Dense cloud  & $0.01$ &  $10^{49}$ & Ionized & $0.3$& $500$ & Dashed line \\
          Dense cloud & $0.1$ &  $10^{50}$ & Neutral & $0.3$& $500$ & Solid line\\
         Dense cloud  & $0.01$ &  $10^{49}$ & Neutral & $0.3$& $500$ & Dashed line 
    \end{tabular}
    \caption{{{ The relevant parameters used throughout the paper. Here we adopt $n=10$, for the outer ejecta index (see Eq.~\ref{eq:Fq} and \(s=2\) as the ambient-density index in the initial Chevalier solution (see Eq.~\ref{eq:rhow} and text thereafter).}} 
    In the Neutral upstream case, we take the external (ambient) and the pre-AIC shell material ionization levels to be neutral, i.e., $x_{\rm e,ext},x_{e,{\rm shell}}^{\rm up}=0,0$, whereas in the ionized-upstream case they are $x_{\rm e,ext},x_{e,{\rm shell}}^{\rm up}=1,1$. A more realistic case follows the ionization-level evolution, \citep[e.g.,][]{Metzger26}.  We discuss the choices of $M_{\rm ejc}$ and the ejecta energy in \S \ref{sec:diss}. }
    \label{tab:modelparamters}
\end{table*}

\subsection{AIC ejecta}\label{sec:ejecta}

We adopt a phenomenological model for the freely expanding AIC ejecta that preserves the total ejecta mass, $M_{\rm ejc}$, and kinetic energy, $E_K$, while allowing the ejecta geometry to be aspherical. Once the initial acceleration has ceased, each fluid element moves ballistically, such that $r=vt$. We write the outer ejecta radius along the direction $\theta$ as $R_{\rm ejc}(\theta,t)=v_{\rm ejc}(\theta)t$ and introduce the dimensionless velocity coordinate
\begin{equation}
    q\equiv\frac{r}{R_{\rm ejc}(\theta,t)}
      =\frac{v}{v_{\rm ejc}(\theta)},
    \qquad 0\leq q\leq 1.
\end{equation}
Surfaces of constant $q$ therefore comove with the homologously expanding ejecta.

The mass contained in the interval $(v,v+dv)$ and solid angle $d\Omega$ is
\begin{equation}
    dM_{\rm ejc}=\rho_{\rm ejc}\,r^2dr\,d\Omega
    =\rho_{\rm ejc}\,v^2t^3dv\,d\Omega.
\end{equation}
Mass conservation then gives
\begin{align}
    \rho_{\rm ejc}(r,\theta,t)
    &=
    \frac{dM_{\rm ejc}}{d\Omega}(\theta)
    \frac{1}{R_{\rm ejc}^3(\theta,t)I_2}
    F\left(\frac{r}{R_{\rm ejc}}\right)
    \nonumber\\
    &=
    \frac{dM_{\rm ejc}}{d\Omega}(\theta)
    \frac{1}{v_{\rm ejc}^3(\theta)t^3I_2}
    F\left(\frac{v}{v_{\rm ejc}}\right)  , 
    \label{eq:rhoejc}
\end{align}
where $I_2=I_{k=2}(0,1)$ is the following integral for $k=2$ and $a=0$ and $b=1$
\begin{equation}
    I_k(a,b)=\int_a^b q^kF(q)\,dq . 
   \label{eq:Ik}
\end{equation}
%
This normalization ensures that
$\int_0^{R_{\rm ejc}}\rho_{\rm ejc}r^2dr=dM_{\rm ejc}/d\Omega$.
The corresponding mass--velocity distribution is
\begin{equation}
    \frac{d^2M_{\rm ejc}}{d\Omega\,dv}
    =
    \frac{dM_{\rm ejc}}{d\Omega}(\theta)
    \frac{q^2F(q)}{v_{\rm ejc}(\theta)I_2}.
    \label{eq:mass_velocity}
\end{equation}

We represent the radial structure of the ejecta with a continuous broken power
law, a standard idealization of freely expanding explosive ejecta
\citep[e.g.,][]{Chevalier82,ChevalierSoker89,TrueloveMcKee99}:
\begin{equation}
    F(q)=
    \begin{cases}
        q^{-\delta}, &
        0\leq q\leq q_{\rm core},\\[3pt]
        q_{\rm core}^{\,n-\delta}q^{-n}, &
        q_{\rm core}<q\leq 1 \ ,
    \end{cases}
    \label{eq:Fq}
\end{equation}
{{where $n$ is the outer ejecta index. }}
This form describes a shallow inner core containing most of the ejecta mass and a steep, low-mass envelope extending to the highest velocities. We adopt $n=10$, $\delta=0$, and $q_{\rm core}=0.1$, consistent with the fiducial structured-ejecta model employed by \citet{Zhao+21}. 
{{Note that $\delta=0$ is both consistent with finite core mass \citep{Chevalier82} and yields a converging DM in the core for a constant ionization fraction \citep{Zhao+21}.  We have verified that varying the power-law index $n$ in the range $[6,14]$ 
produces qualitatively similar results, and the external densities and the upstream ionization fraction yield larger qualitative DM differences.  }}

We introduce asphericity by assuming that surfaces of constant $q$ are concentric, homologously expanding, similar spheroids. Denoting the polar and equatorial velocities of the outermost ejecta by $v_{\rm pol}$ and $v_{\rm eq}$, respectively, the outer velocity in the direction $\theta$ is
\begin{equation}
    v_{\rm ejc}(\theta)
    =
    \left(
        \frac{\sin^2\theta}{v_{\rm eq}^2}
        +
        \frac{\cos^2\theta}{v_{\rm pol}^2}
    \right)^{-1/2}.
    \label{eq:vejc}
\end{equation}
Requiring the density to be constant on each spheroidal surface of constant
$q$ implies
\begin{equation}
    \frac{dM_{\rm ejc}}{d\Omega}(\theta)
    =
    \frac{M_{\rm ejc}\,v_{\rm ejc}^3(\theta)}
         {4\pi v_{\rm pol}v_{\rm eq}^2},
    \label{eq:dMdOmega}
\end{equation}
where we have used
$\int v_{\rm ejc}^3(\theta)\,d\Omega
=4\pi v_{\rm pol}v_{\rm eq}^2$.
Consequently, a prolate configuration (like the ejecta propagation in the presence of an equatorial wind and a jet) contains more mass per unit solid angle near the polar axis than near the
equator. In other words: $\mathcal{A}\equiv{v_{\rm pol}}/{v_{\rm eq}}>1$.
The total kinetic energy is
\begin{equation}
    E_K
    =
    \frac{1}{2}\frac{I_4}{I_2}
    \int d\Omega\,
    \frac{dM_{\rm ejc}}{d\Omega}(\theta)
    v_{\rm ejc}^2(\theta) \ .
    \label{eq:Ek}
\end{equation}
{{Note that the factor $I_4/I_2$ arises because the mass integral is weighted by $q^2F(q) dq$, whereas the kinetic-energy integral includes the additional factor $v^2=q^2v_{\rm ejc}^2(\theta)$ and therefore involves $q^4F(q)\,dq$.
Thus, $I_4/I_2$ is the mass-weighted mean of $q^2$, converting the squared outer-edge velocity into the mean squared velocity within each angular sector.
}}
For a spheroid, {{the integral that one gets when plugging Equation~(\ref{eq:dMdOmega}), into Equation~(\ref{eq:Ek}) is: }}
\begin{equation}
    \int v_{\rm ejc}^5(\theta)\,d\Omega
    =
    \frac{4\pi}{3}v_{\rm pol}v_{\rm eq}^2
    \left(2v_{\rm eq}^2+v_{\rm pol}^2\right).
\end{equation}
The equatorial velocity is therefore fixed by
$(M_{\rm ejc},E_K,\mathcal{A})$:
\begin{equation}
    v_{\rm eq}^2
    =
    \frac{6E_K I_2}
         {M_{\rm ejc}I_4\left(2+\mathcal{A}^2\right)},
    \qquad
    v_{\rm pol}=\mathcal{A}v_{\rm eq} \ ,
    \label{eq:veq}
\end{equation}
{{with $\mathcal{A}=1.5$. This value corresponds to a polar outer-edge velocity $50\%$ greater than the equatorial value. This prolate prescription is motivated qualitatively by multidimensional AIC simulations showing preferential polar ejection associated with rotation and, in magnetized models, magnetic stresses \citep[e.g.,][]{Dessart+06,Dessart+07}. }}

In the steep outer envelope, $q>q_{\rm core}$, Equation~(\ref{eq:rhoejc})
can be written as
\begin{equation}
    \rho_{\rm ejc}(r,\theta,t)
    =
    g_n(\theta)t^{\,n-3}r^{-n},
    \label{eq:gn}
\end{equation}
where
\begin{equation}
    g_n(\theta)
    =
    \frac{dM_{\rm ejc}}{d\Omega}(\theta)
    \frac{q_{\rm core}^{\,n-\delta}
          v_{\rm ejc}^{\,n-3}(\theta)}
         {I_2}.
    \label{eq:gn_norm}
\end{equation}
Thus, within each angular sector, the outer ejecta have the same radial and
temporal scaling as the self-similar ejecta profile of \citet{Chevalier82}.
The quantity $g_n(\theta)$ is the direction-dependent analogue of the normalization denoted by $g^n$ in that work.

Throughout the paper, we examine two cases for the ejecta parameters (see Table \ref{tab:modelparamters}), specifically: \((M_{\rm ejc}, E_{\rm k})=(0.01~{\rm M}_\odot,10^{49} {\rm erg})\) and \((0.1 ~ {\rm M}_\odot,10^{50} {\rm erg})\) as representative weak and energetic cases, respectively, \citep[e.g.,][]{Dessart+06,Dessart+07,Metzger+09,Darbha+10,Yu+19}. In both cases, the bulk ejecta remain subrelativistic. 

We note that at early times, while the interaction retains memory of the homologous ejecta profile, we use the ejecta-dominated self-similar solution of \citet{Chevalier82}. The Sedov--Taylor regime is instead the late-time asymptotic limit in which the swept-up external mass dominates, and the explosion energy is approximately conserved; in our calculation this limit is not imposed separately but is contained in the subsequent energy-conserving shell evolution.

\section{DM components }\label{sec:zones}
\begin{figure*}
    \centering
    \includegraphics[width=1\linewidth]{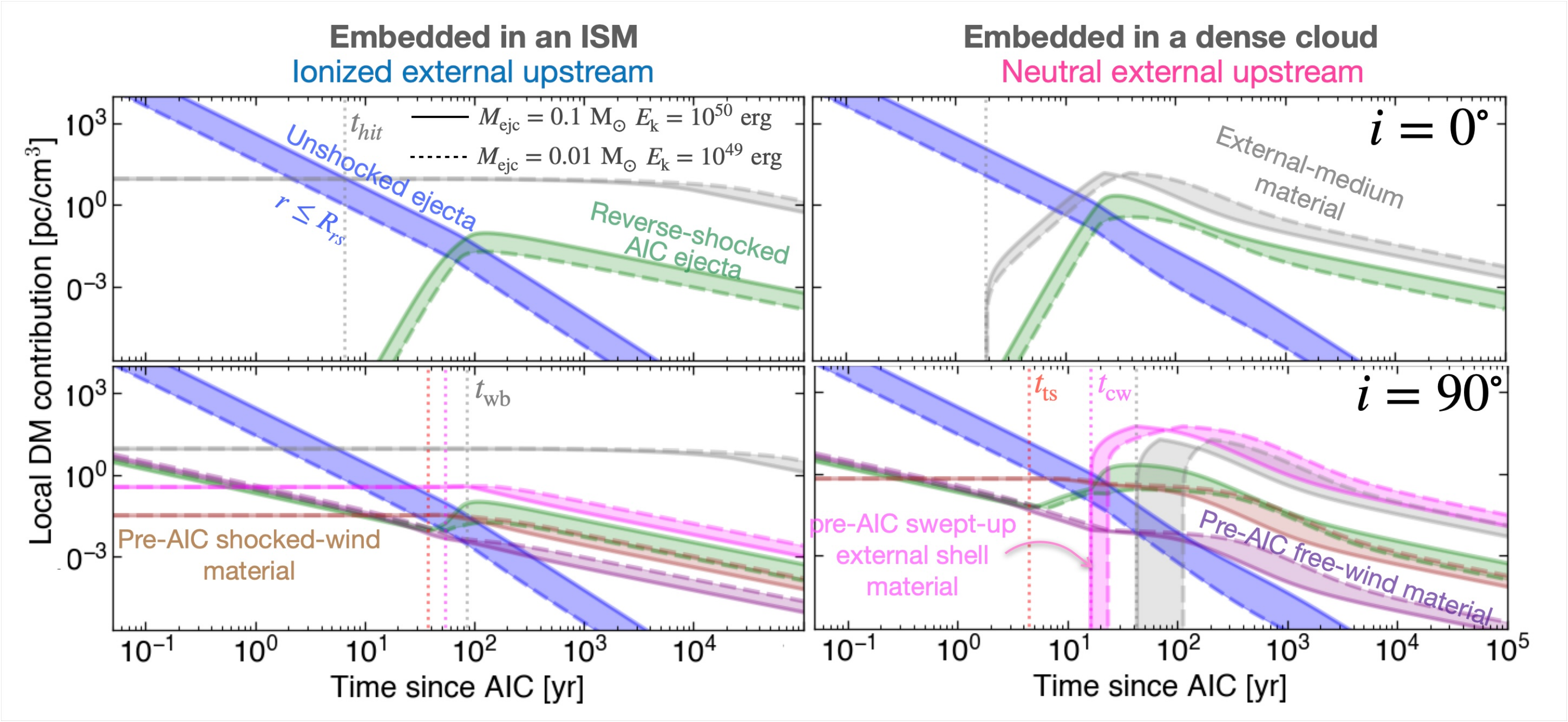}
    \caption{ Contributions to the local DM, grouped by material origin, as a function of time since the AIC. The rows show viewing inclinations $i=0^\circ$ and $90^\circ$. The left column shows an ionized external upstream embedded in a diffuse ISM with $n_{\rm ext}=1~{\rm cm^{-3}}$ and $R_{\rm out}=10~{\rm pc}$, while the right column shows a system embedded in a dense cloud with $n_{\rm ext}=500~{\rm cm^{-3}}$ and $R_{\rm out}=0.3~{\rm pc}$. In the dense-cloud example, the pre-AIC swept-up external shell and the external medium ahead of the AIC shock are neutral, whereas the AIC shock ionizes the processed material. The components are the unshocked AIC ejecta (blue), reverse-shocked AIC ejecta (green), pre-AIC free-wind material (purple), pre-AIC shocked-wind material (brown), pre-AIC swept-up external-medium shell material (magenta), and external-medium material (gray). Each circumstellar or external-medium component includes both the material already swept into the AIC-shocked shell and any material that remains ahead of the AIC forward shock. Solid and dashed curves correspond to $(M_{\rm ejc},E_K)=(0.1\,M_\odot,10^{50}\,{\rm erg})$ and $(0.01\,M_\odot,10^{49}\,{\rm erg})$, respectively; the shaded bands span these two ejecta models. The vertical dotted lines mark the impact with the polar cavity wall, $t_{\rm hit}$; crossings of the pre-AIC wind termination shock, contact discontinuity, and outer bubble shock, $t_{\rm ts}$, $t_{\rm cw}$, and $t_{\rm wb}$. 
    These vertical landmarks correspond to the solid-curve ejecta model. By construction, the pre-AIC wind and swept-up shell components vanish in the polar direction. }
    \label{fig:Dynamics}
\end{figure*}

{We now compute the local DM along a sightline at inclination $i$ by summing the contributions from each zone of the structure described in Section~\ref{sec:ToyModel}, 
\begin{equation} {\rm DM}_k(i,t) = \int_{R_1}^{R_2} n_e(r,\theta=i,t)\,dr , \label{eq:DMk} \end{equation} 
where $n_e$ is given by Equation~(\ref{eq:ne}) and the interface radii $R_1$ and $R_2$ follow from the dynamical prescription in Appendices~\ref{app:dynamics} and~\ref{app:thin_shell}. We group the contributions by the origin of the material from the inside out: the unshocked and reverse-shocked AIC ejecta (\S \ref{sec:unshejec}--\ref{sec:shejec}), the pre-AIC free wind (\S \ref{sec:fw}), the pre-AIC shocked wind (\S \ref{sec:sw}), the pre-AIC swept-up external shell (\S \ref{sec:shell}), and the external medium (\S \ref{sec:ext}). Each circumstellar component includes both the material still ahead of the AIC forward shock and the material already swept into the AIC-shocked shell. Any contribution whose upper radial boundary does not exceed its lower boundary is set to zero. 

Figure~\ref{fig:Dynamics} shows these components as a function of time since the AIC for inclinations $i=0^\circ$ 
and $90^\circ$, for two of the environments listed in Table~\ref{tab:modelparamters}: a diffuse medium ($n_{\rm ext}=1\,{\rm cm^{-3}}$) with ionized upstream gas (left), and a dense cloud ($n_{\rm ext}=500\,{\rm cm^{-3}}$) with neutral upstream gas (right). Solid and dashed curves bracket the two ejecta models. Because time is measured from the AIC, the results are agnostic to when FRB activity begins. The total DM and its time derivative are discussed in Section~\ref{sec:DMvar}. 

The viewing-angle dependence is immediately apparent. A polar sightline ($i=0^\circ$) passes through the evacuated cavity and therefore misses the wind components entirely, producing larger DM variability; the same source can thus exhibit quite different DM histories depending on our viewing angle.
}

    \subsection{Unshocked ejecta ($0\le r<R_{\rm rs}$)}\label{sec:unshejec} 
    In this component, we integrate Equation (\ref{eq:DMk}) with the density presented in Equation~(\ref{eq:rhoejc}) with $dr=v_{\rm ejc}t\,dq$,
\begin{equation}\label{eq:DMunshejc}
  {\rm DM}_{\rm unsh,ejc}=\frac{x_{e,{\rm ejc}}Y_{e,{\rm ejc}}}{m_p}\,
  \frac{dM_{\rm ejc}/d\Omega}{I_2\,v_{\rm ejc}^2\,t^2}\int_0^{q_{\rm rs}}F(q)\,dq\ ,
\end{equation}
where $I_2=I_{k=2}$ from Equation (\ref{eq:Ik}), 
\begin{equation}
    q_{\rm rs}\equiv\min\!\left(\frac{R_{\rm rs}}{v_{\rm ejc}t},1\right) \ ,
\end{equation}
and $\int_0^{q}F\,dq=q^{1-\delta}/(1-\delta)$ for $q\le q_{\rm core}$ and
$q_{\rm core}^{1-\delta}/(1-\delta)+q_{\rm core}^{\,n-\delta}(q_{\rm core}^{1-n}-q^{1-n})/(n-1)$, otherwise.  Before core entry, the integral is dominated by the core, and ${\rm DM}_{\rm unsh,ejc}\propto t^{-2}$. 
Here each element keeps its mass, and its radius grows as $\propto t$. After core entry, $I_{k=0}(0,q_{\rm rs})\to R_{\rm rs}/(v_{\rm ejc} t)$, see Eq.~(\ref{eq:Ik}); thus, the DM steepens such that $\propto t^{-3}$. 
In other words:
\begin{equation}
    DM_{\rm unsh,ejc}\propto
    \begin{cases}
        t^{-2}, &
        t\leq t_{\rm core},\\[3pt]
        t^{-3} , &
        t> t_{\rm core} \ .
    \end{cases}
    \label{eq:scalingDMsunchejc}
\end{equation}
Figure \ref{fig:Dynamics} shows this component as the dark blue band.


\subsection{Reverse-shocked AIC ejecta ($R_{\rm rs}\le r<R_{\rm cd}$)}\label{sec:shejec}
The shocked ejecta comprise all ejecta with $q>q_{\rm rs}$,
\begin{equation}
  \frac{dM_{\rm sh,ejc}}{d\Omega}=\frac{dM_{\rm ejc}}{d\Omega}\,
  \frac{\int_{q_{\rm rs}}^1q^2F\,dq}{I_2} \ ,
   \label{eq:dmB}
\end{equation}
\begin{equation}\label{eq:DMshejc}
     {\rm DM}_{\rm sh,ejc}=\frac{x_{e,{\rm ejc}}Y_{e,{\rm ejc}}}{m_p}\,
  \frac{3\,dM_{\rm sh,ejc}/d\Omega}{R_{\rm cd}^2+R_{\rm cd}R_{\rm rs}+R_{\rm rs}^2} \ .
\end{equation}
Further, ${\rm DM}_{\rm sh,ejc}=0$ in the polar cone for $t\le t_{\rm hit}$, where  $t_{\rm hit}$ is the time at which the polar ejecta reach the cavity wall. For $t<t_{\rm hit}$, no shocks form, and all ejecta remain unshocked. At $t_{\rm hit}$, a forward shock is launched into the ambient ISM. In the polar cone, $q_{\rm rs}=1$ for $t\le t_{\rm hit}$. Figure \ref{fig:Dynamics} shows this component as a light green band. 

The slope of the DM, is understood from the above equations. In the envelope: ${dM_{\rm sh,ejc}}/{d\Omega}\propto q_{\rm rs}^{3-n}\sim t^{(3-n)(m_s-1)}$, and since $q_{\rm rs}=R_{\rm rs}/(v_{\rm ejc} t)\sim t^{m_s-1}$; see Equations (\ref{eq:Rcd_early})-(\ref{eq:tts_appendix}). Plugging this into Equation (\ref{eq:DMshejc}), we have DM$\propto t^{(3-n)(m_s-1)-2m_s}$, which gives a decaying slope for the wind case ($s=2$) and a positive slope for the ambient material ($s=0$).
 In other words: $ DM_{\rm sh,ejc}\sim t^{(3-n)/(n-2)}\sim t^{-7/8}$, for $s=2$ and $ \theta > \theta_{\rm cav}$. For $s=0$ and $\theta \leq \theta_{\rm cav}$, we use our thin-shell calculation -- see Appendix \ref{app:thin_shell}. Nonetheless, we can estimate the expected exponent dependence if one assumes self-similar behavior, which gives:  $DM_{\rm sh,ejc}\propto t^{n-3} \sim t^7$ for $s=0;~ \theta > \theta_{\rm cav}$ and $DM_{\rm sh,ejc}\propto t^{n-3} \sim t^7$ for $\theta \leq \theta_{\rm cav}$. The thin-shell calculation produces some deviations from these power laws, but the qualitative behavior remains. 
Here, the rise takes place for the $s=0$ case, while the decline takes place during the interaction with the wind, $s=2$, as seen in the green band left panels of Figure \ref{fig:Dynamics}. The green slope in the dense-cloud scenario deviates slightly from this scaling because of post-transition thin-shell evolution; see Appendix \ref{app:thin_shell}.

\subsection{Pre-AIC free-wind material}\label{sec:fw}

This component includes all material that initially belonged to the
freely expanding pre-AIC wind,
$R_{\rm w,in}\leq r<R_{\rm ts}(\theta)$. At a given time, this material is divided into two parts: material that remains ahead of the AIC forward shock and material that has
been swept into the AIC-shocked circumstellar shell. We calculate
the two contributions separately and then add them.

The portion of the free wind that has not yet been reached by the AIC forward shock occupies $\max\!\left(R_{\rm fs},R_{\rm w,in}\right)
    \leq r<R_{\rm ts}(\theta)$.
Because  $\rho_w(r,\theta)=K(\theta)r^{-2}$, its DM contribution is
\begin{eqnarray}
     {\rm DM}_{{\rm fw},{\rm ahead}} &=& \frac{x_{e,w}^{\rm up}Y_{e,w}}{m_p} K(\theta) \\ 
    &\times&  \left(   \frac{1}{\max\!\left(R_{\rm fs},R_{\rm w,in}\right)}  -  \frac{1}{R_{\rm ts}(\theta)}   \right)  \ . \nonumber
\end{eqnarray}
 This contribution vanishes once $R_{\rm fs}\geq R_{\rm ts}$.

The free-wind material already encountered by the AIC forward shock originally occupied $R_{\rm w,in} \leq r < \min\!\left(R_{\rm fs},R_{\rm ts}\right)$.  Its swept-up mass per unit solid angle is
\begin{equation}
    \frac{dM_{{\rm fs},{\rm AIC}}}{d\Omega}   = K(\theta)  \left(     \min\!\left(R_{\rm fs},R_{\rm ts}\right) -R_{\rm w,in} \right)  \ .
    \label{eq:free_wind_swept_mass}
\end{equation}
We approximate this material as being distributed uniformly in the
current AIC-shocked circumstellar shell,
$R_{\rm cd}\leq r<R_{\rm fs}$. Its DM contribution is therefore
\begin{equation}
    {\rm DM}_{{\rm fw},{\rm AIC}}
    =
    \frac{x_{e,w}^{\rm down}Y_{e,w}}{m_p}
    \frac{
        3\,dM_{{\rm fw},{\rm AIC}}/d\Omega
    }{
        R_{\rm fs}^2
        +R_{\rm fs}R_{\rm cd}
        +R_{\rm cd}^2
    } .
    \label{eq:dm_free_wind_swept}
\end{equation}

The total DM associated with pre-AIC free-wind material is thus
\begin{equation}
    {\rm DM}_{\rm fw}
    =
    {\rm DM}_{{\rm fw},{\rm ahead}}
    +
    {\rm DM}_{{\rm fw},{\rm AIC}} .
    \label{eq:dm_free_wind_total}
\end{equation}
No pre-AIC free-wind material is present inside the polar cavity, so
both contributions vanish in polar directions. Figure
\ref{fig:Dynamics} shows the total pre-AIC free-wind contribution as
the purple band.

Inside the shocked shell sectors, after a finite reservoir has been swept, the DM is thus proportional to: 
\begin{equation}
{\rm DM}_{j,\rm sh}\propto
\left(R_{\rm fs}^{2}+R_{\rm fs}R_{\rm cd}+R_{\rm cd}^{2}\right)^{-1}
\sim R_{\rm fs}^{-2} \ ,
\end{equation}
where $j$ stands for shocked sectors. 



\subsection{Pre-AIC shocked-wind material}\label{sec:sw}
This component is associated with material that belonged to the pre-AIC shocked-wind bubble. Specifically, we identify two regimes. In the first, we have DM from pre-AIC shocked-wind material swept into the AIC shocked-circumstellar matter shell. Here the original material is placed at: $R_{\rm ts} \leq r < {\rm min}(R_{\rm fs}, R_{\rm cw})$ and the shell is approximated to be at: $R_{\rm cd}\le r<R_{\rm fs}$. The material swept here is:
\begin{equation}
    \frac{dM_{\rm sw,sh}}{d\Omega}=\frac{\rho_{\rm sw}(\theta)}{3}\left[\min(R_{\rm fs},R_{\rm cw})^3-R_{\rm ts}^3\right] \ .
\end{equation}
The other regime is associated with the pre-AIC shocked wind that remains ahead of the AIC forward shock. This regime spans ${\rm max}(R_{\rm fs}, R_{\rm ts}) \leq r < R_{\rm cw}$. The material here is: 
    \begin{equation}
  {\rm DM}_{\rm sw,ahead}=\frac{x_{e,sw}Y_{e,w}}{m_p}\,\rho_{\rm sw}(\theta)
  \left[R_{\rm cw}-\max(R_{\rm fs},R_{\rm ts})\right] \ .
  \label{eq:dmahead}
\end{equation}
Additionally, this component also includes the swept shocked-wind material, i.e.,
\begin{equation}
    {\rm DM}_{\rm sw,sh} = \frac{x_{e,\rm sw}^{\rm down}Y_{e,\rm w}}{m_p}
\frac{3\,dM_{\rm sw,sh}/d\Omega} {R_{\rm fs}^{2}+R_{\rm fs}R_{\rm cd}+R_{\rm cd}^{2}} \ .
\end{equation}
Thus, overall this component has: 
\begin{equation}
  {\rm DM}_{\rm sw} = {\rm DM}_{\rm sw,sh}+{\rm DM}_{\rm sw,ahead} \ .
\end{equation}

  This component vanishes in the polar cone. 
  Figure \ref{fig:Dynamics} depicts this component as the light-brown band. 

\subsection{ pre-AIC swept-up external shell material}\label{sec:shell}
This component describes the DM from material belonging to the pre-AIC swept-up external material shell. The first regime is associated with the pre-AIC swept-up external material that is subsequently swept again by the AIC forward shock; this material originated from $R_{\rm cw}\leq r < {\rm min}(R_{\rm fw},R_{\rm wb})$ and we approximate the shell to be distributed uniformly between $R_{\rm cd}\leq r < R_{\rm fs}$. 
The other regime belongs to the part of the pre-AIC swept-up external shell
    that remains ahead of the AIC forward shock. This region spans  ${\rm max}(R_{\rm fs},R_{\rm cw})\leq r < R_{\rm wb}$. In the neutral upstream case, this regime's ionization level $x_{e,\rm shell}^{\rm up}\rightarrow 0$.
  The mass per unit solid angle
from the pre-AIC swept-up shell that has subsequently been swept
into the AIC-shocked shell is
\begin{equation}
    \frac{dM_{shell,{\rm sh}}}{d\Omega}
    =
    \frac{\rho_{\rm shell}}{3}
    \left[
        \min\!\left(R_{\rm fs},R_{\rm wb}\right)^3
        -R_{\rm cw}^3
    \right] \ ,
    \label{eq:pre_shell_reswept_mass}
\end{equation}
where $\rho_{\rm shell}$ is defined in Equation (\ref{eq:rhosh}). 
We distribute this material uniformly between $R_{\rm cd}$ and $R_{\rm fs}$, giving
\begin{equation}
    {\rm DM}_{\rm shell, sh}     =
    \frac{x_{e,{\rm shell}}^{\rm down}Y_{e,{\rm ext}}}{m_p} \frac{ 3\,dM_{\rm shell,{\rm sh}}/d\Omega }{ R_{\rm fs}^2 +R_{\rm fs}R_{\rm cd}   +R_{\rm cd}^2 } \ .
    \label{eq:dm_pre_shell_reswept}
\end{equation}

The portion of the pre-AIC swept-up shell that remains ahead of
the AIC forward shock occupies
$\max(R_{\rm fs},R_{\rm cw})\leq r<R_{\rm wb}$ and contributes
\begin{equation}
    {\rm DM}_{\rm shell,up} =\frac{x_{e,{\rm shell}}^{\rm up}Y_{e,{\rm ext}}}{m_p} \rho_{\rm shell} \left[  R_{\rm wb}  -\max\!\left(R_{\rm fs},R_{\rm cw}\right)
    \right] \ .
    \label{eq:dm_pre_shell_ahead}
\end{equation}
In the upstream neutral case, $x_{e,{\rm shell}}^{\rm up}\to 0$. 
Figure \ref{fig:Dynamics} depicts this component as the magenta band. 
\subsection{External-medium material}\label{sec:ext}
Here, ``external-medium material'' denotes material initially contained within the explicitly modeled external reservoir, \(R_{\rm ext,in}(\theta)\le r<R_{\rm out}\). In the dense-cloud calculation, the dynamics continue into an \(n_{\rm bg}=1~{\rm cm^{-3}}\) background after the forward shock passes \(R_{\rm out}\), but the DM associated with this background is not included.
This component also includes two regimes. The first is external material swept by the AIC forward shock, and thus shocked.  We assume the original ambient-ISM region is at $R_{\rm cav}$ in the polar direction and $R_{\rm wb}$ in nonpolar directions. We include only material initially located between this inner boundary $R_{\rm out}$. 
Thus, we define:
\begin{equation}
    R_{\rm ext,in}(\theta) = \left\{ \begin{array}{ll} 
    R_{\rm cav} \ , \theta\leq \theta_{\rm cav} \\ 
    R_{\rm wb} \ , \theta >  \theta_{\rm cav} \nonumber
    \end{array} \right.
\end{equation}
The swept material is placed uniformly in the $R_{\rm cd}\leq  R_{\rm fs}$ shell
\begin{equation}
      \frac{dM_{{\rm ext,sh}}}{d\Omega}=\frac{\rho_{\rm ext}}{3}\left[\min(R_{\rm fs},R_{\rm out})^3-R_{\rm ext,in}^3\right] \ ,
  \label{eq:MCism}
\end{equation}
and the corresponding DM is:
\begin{equation}
  {\rm DM}_{\rm ext,sh}=\frac{x_{e}Y_{e}}{m_p}\,
  \frac{3\,dM_{{\rm ext,sh}}/d\Omega}{R_{\rm fs}^2+R_{\rm fs}R_{\rm cd}+R_{\rm cd}^2} . \
  \label{eq:dmswept}
\end{equation}
The second regime involves the unshocked material and spans over:  ${\rm max}[R_{\rm fs}(t, \theta), R_{\rm ext,in}(\theta)] \leq r < R_{\rm out}$,
\begin{eqnarray}
     {\rm DM}_{\rm ext,ahead}&=&x_{e,ext}^{up} Y_{e,ext}n_{\rm ext} \\ &\times& \left[R_{\rm out}-{\rm max}(R_{\rm fs},R_{\rm ext,in})\right] \ . \nonumber \label{eq:dmsweptahead}
\end{eqnarray}
In the neutral upstream scenario, $x_{e,{\rm ext}}\to 0$. Figure \ref{fig:Dynamics} depicts this component as the gray band. 

\begin{figure*}
    \centering
    \includegraphics[width=1\linewidth]{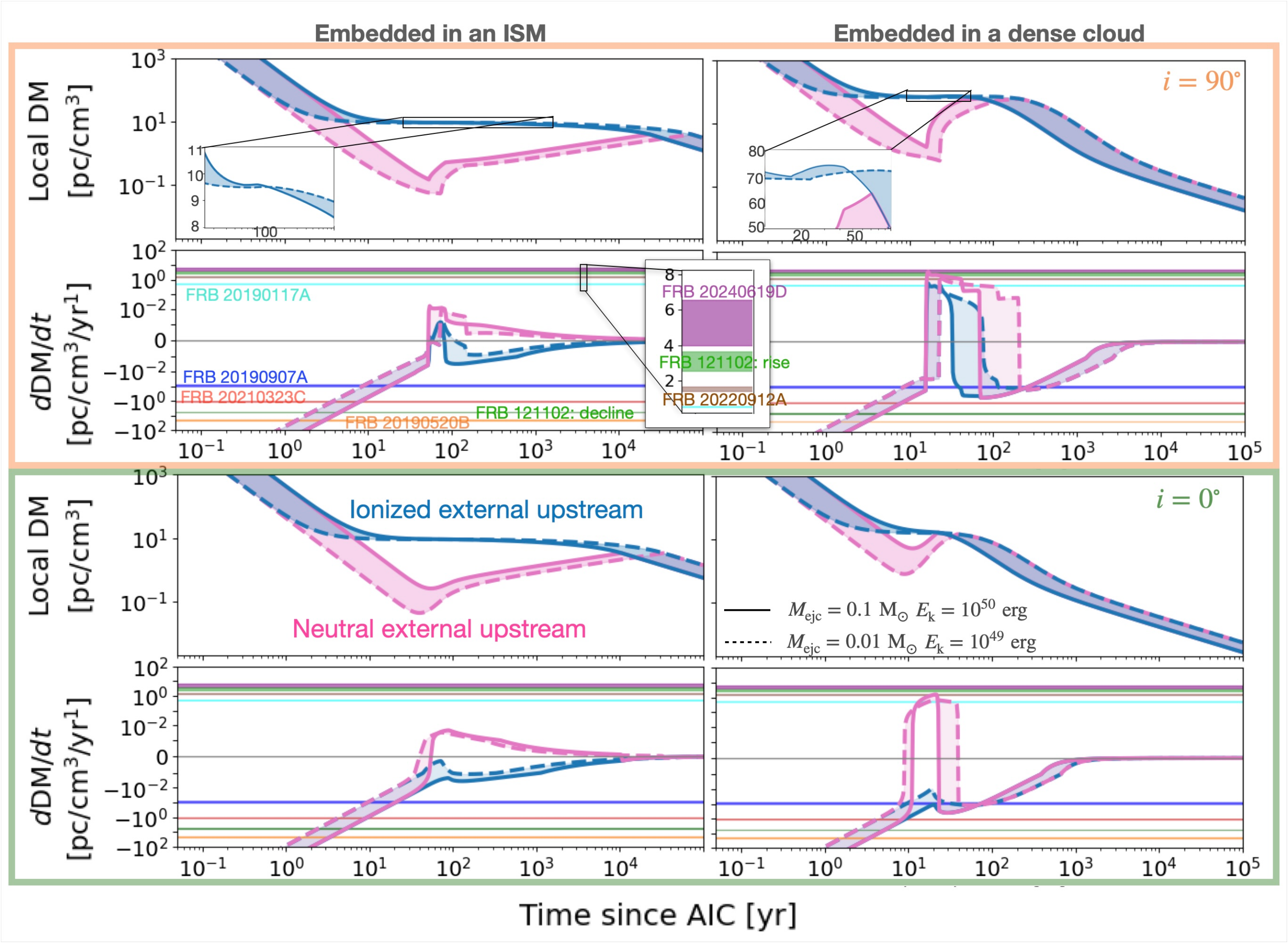} 
    \caption{Evolution and variability of the total local DM for
an equatorial sightline, $i=90^\circ$, top, and $i=0^\circ$, bottom. The left and right columns show systems embedded in a diffuse ISM
($n_{\rm ext}=1~{\rm cm^{-3}}$, $R_{\rm out}=10~{\rm pc}$) and a dense cloud ($n_{\rm ext}=500~{\rm cm^{-3}}$, $R_{\rm out}=0.3~{\rm pc}$), respectively. Blue curves show the ionized-upstream model, while magenta curves show the neutral-upstream prescription, in which the pre-AIC swept-up external shell and the unshocked external medium are neutral
but become ionized after being processed by the AIC shock. The top row shows the total local DM, and the bottom row shows its time derivative, $d{\rm DM}_{\rm loc}/dt$, with the horizontal gray line marking zero. The labeled horizontal colored lines indicate measured positive and
negative DM-variation rates for selected repeating FRBs and are shown for comparison only; the models have not been fitted to the individual sources. 
 Specifically, values for FRB 20190117A, FRB 20190907A, FRB 20210323C are estimated from \citet{Cook+26}, and FRB 121102, FRB 20240619D, FRB 20220912A and  FRB 20190520B are  estimated from \citet{Hessels2019,Ould-Boukattine+26,Abbott+26FRB} and \citet{Niu+26}, respectively. We split FRB 121102 into the rising and declining branches.  
 The observed DM variations are in the rest frame of the source, i.e. have been corrected for a $(1+z)^2$ factor.
For FRB 20240619D we used the upper limit $z\leq 0.37$ from \citet{Ould-Boukattine+26}, while those of FRBs 20190117A and 20190907A are the maxima inferred from their DMs, 0.46 and 0.33,
respectively \citep{Michilli+23}. FRB 20220912A's $z=0.077$ is from \citet{Ravi+23}, FRB 20190520B's $z=0.24$ is from \citet{Niu2022} and FRB 121102's $z=0.19$ is from \citet{Tendulkar+17}. For FRB 20210323C, there is no redshift yet known, so we assume $z=0.2$ given its observed DM, and estimates for the Milky Way and host contributions \citep{Macquart+20,shin+2023,ocker+2026}.   
Solid and dashed curves correspond to
$(M_{\rm ejc},E_K)=(0.1\,M_\odot,10^{50}\,{\rm erg})$ and
$(0.01\,M_\odot,10^{49}\,{\rm erg})$, respectively; the shaded regions span the two ejecta models. The structural transitions produce robust changes in the DM slope, whereas the sign and possible reversal of the total DM evolution depend on the ionization prescription and external environment. The top inset zooms in to highlight the structure of the ionized external upstream case. {{Note that while it seems that the slope reaches zero at long timescales, it's an artifact of the log scale and it remains negative, but shallow.  }}}
    \label{fig:Compared2observations}
\end{figure*}


\section{DM variability}\label{sec:DMvar}

We explore DM variability as a function of time within the toy-model framework by considering $d{\rm DM}/dt$. Specifically, Figure \ref{fig:Compared2observations} shows the total DM calculated according to \S \ref{sec:zones} (top panels) and its time derivative (bottom panels). 
Figure \ref{fig:Compared2observations} depicts the DM time evolution for $i=90^\circ$, top, and $i=0^\circ$ bottom. 

For the fiducial models shown in Figures~\ref{fig:Dynamics} and \ref{fig:Compared2observations}, the transitions occur between a few years and approximately a century after AIC.
{{The polar sightline bypasses the pre-AIC wind zones and first encounters matter when the ejecta strike the cavity wall at \(t_{\rm hit}\simeq1.9~{\rm yr}\) (Figure~3). Along nonpolar sightlines, the forward shock instead reaches \(R_{\rm ts}\), \(R_{\rm cw}\), and \(R_{\rm wb}\) in succession.}} 
On the other hand, in the equatorial line of sight, the forward shock reaches the wind termination shock at $t_{\rm ts}\simeq2.9$--$6.1$~yr, the wind-bubble contact discontinuity at $t_{\rm cw}\simeq12$--$23$~yr, and the outer wind-bubble shock at $t_{\rm wb}\simeq31$--$113$~yr, spanning the two ejecta models. 
In the initially neutral models, the first negative-to-positive crossing of $d{\rm DM}_{\rm loc}/dt$ occurs at $t\simeq9$--$22$~yr, when the rate at which the shock converts neutral material into free electrons overtakes the declining contribution from the expanding ejecta. In the dense-cloud model, the ensuing rising branch reaches sustained rates of order a few, approximately $3$--$5~{\rm pc\,cm^{-3}\,yr^{-1}}$, away from the narrow excursions associated with the idealized sharp interfaces. These values are illustrative rather than fits to individual sources, but they demonstrate that the structural transitions and DM reversals naturally occur on the decade-to-century timescales probed by observations of repeating FRBs.

The left panels of Figure~\ref{fig:Compared2observations} 
depict a neutral (ionized) upstream source embedded in a diffuse external medium (i.e., $n_{\rm ext}=1$~cm$^{-3}$), with a pink (blue) shaded region. To the right, we consider a source embedded in a dense cold cloud (i.e., $n_{\rm ext}=500$~cm$^{-3}$). To explore all permutations, we bracket the ionization levels here between ionized-upstream at the source (blue) and the neutral upstream case (pink).  For both cases, we adopt two limiting cases for the ejecta and kinetic energies (solid and dashed lines that bracket the shaded regions). 

In the $d{\rm DM}/dt$ panel of the Figure, we also compare to a few observations. We stress that we do not intend to match or fit any particular source, but to qualitatively demonstrate which source elements produce the observed variability.  We compare to the estimated $d{\rm DM}/dt$ of a few observations. Specifically, FRB 20190117A, FRB 20190907A, FRB 20210323C are estimated from \citet{Cook+26}, and FRB 20121102A (also known as FRB 121102), FRB 20240619D, FRB 20220912A and  FRB 20190520B  estimated from \citet{Hessels2019,Ould-Boukattine+26,Abbott+26FRB} and \citet{Niu+26}, respectively. We particularly split FRB 121102 into the rising and declining branches.


As seen in Figure \ref{fig:Compared2observations},  
large secular rates favor a dynamically young environment, but the inferred age is degenerate with ionization, density structure, and viewing angle. The neutral-upstream prescription produces the largest and most sustained positive DM-variation rates, although smaller positive excursions can also accompany structural transitions in an already ionized medium.



\section{Discussion and Conclusions}\label{sec:diss}

 Secular changes in the DM of repeating FRBs provide a direct probe of the plasma local to their engines. The observed diversity of the rate of change of DM as a function of time, including both positive and negative slopes \citep[e.g.,][]{Hessels2019,Niu2022,WangP2025,Snelders+25,Cook+26,Niu+26,Jia+26,Pandhi2026}, motivates models that treat the ejecta and the circumsource medium together \citep[e.g.,][]{Piro16,Zhao+21,Waxman+26,Metzger26}. AIC of a massive white dwarf provides a delayed pathway to forming an FRB-emitting neutron star \citep[e.g.,][]{Totani13,Shariat+26}. Because AIC is the endpoint of binary mass transfer or white-dwarf merger, the surrounding medium can have an anisotropic circumstellar structure consisting of a freely expanding wind, a wind termination shock, a shocked-wind bubble, a swept-up ambient shell, and the surrounding interstellar medium; see the illustration in Figure \ref{fig:Cartoon}.

Here we demonstrated that an AIC event inside a finite, binary-sculpted circumstellar structure does not produce a universal power-law DM history. Instead, the successive interface crossings, external material density, upstream ionization, and viewing angle determine whether the local DM declines, rises, or reverses trend. We present two scenarios for the external material density: one with the source embedded in $1$~cm$^{-3}$ at $10$~pc and the other in $500$~cm$^{-3}$ at $0.3$~pc. We choose both such that the external environment lies beyond $R_{\rm wb}$. These scenarios serve as proof of environmental effects rather than specific predictions for FRB environments. 


\subsection{The sign and shape of the DM evolution}
Our results suggest that the local DM may have complex, sign-changing secular time evolution when the collapse occurs inside a finite, binary-sculpted circumstellar structure.  At early times, the electron column density is dominated by the freely expanding AIC ejecta and decreases approximately as ${\rm DM}_{\rm ejc}\propto t^{-2}$ (see Figures \ref{fig:Dynamics} and \ref{fig:Compared2observations}, as well as \S \ref{sec:unshejec}). At later times, {if not seen along the pole,} the forward shock successively encounters the wind termination shock, the shocked-wind bubble, the pre-AIC swept-up shell, and the external medium. Each crossing changes both the rate at which mass is swept up and the radius at which that mass contributes to the electron column. The resulting breaks in the DM slope are therefore a generic consequence of the finite circumstellar structure, although their precise sharpness is specific to our piecewise toy-model prescription.

The sign of the DM evolution is controlled primarily by the ionization state of the material ahead of the AIC shock (see Figures \ref{fig:Dynamics} and \ref{fig:Compared2observations}). If the upstream medium is already ionized, passage of the shock mainly redistributes electrons that already contributed to the line-of-sight column density. Moving those electrons into an expanding shell generally reduces their column density, and the total DM consequently declines or passes through a relatively flat phase. If the upstream material is initially neutral, the AIC shock both sweeps up and ionizes new material. The increasing number of free electrons can then overcome the declining ejecta contribution, producing a DM minimum followed by a rise. Once the available neutral reservoir has been processed and the shocked material continues to expand, the DM turns over and declines again. Thus, a rising or
non-monotonic DM does not by itself require an increasing total gas column; it can instead mark the conversion of neutral material into an observable electron column \citep[e.g.][]{Metzger26}. 


\subsection{Viewing angle and the interpretation of an individual source}
The viewing angle with which we observe the source changes the sequence of environments we see the AIC ejecta encounter, so that we do not observe merely a rescaling of the DM. Along nonpolar directions, the forward shock crosses the freely expanding wind, the shocked-wind bubble, and the pre-AIC swept-up shell before reaching the
external medium. Along the evacuated polar channel, these components are absent, and the ejecta expand freely until they strike the cavity wall (as highlighted in Figure \ref{fig:Dynamics}). Consequently, the same intrinsic system can exhibit different transition times, different instantaneous values of $d{\rm DM}/dt$, and
even different signs of the DM evolution depending on line of sight (compare the top and bottom in Figure \ref{fig:Compared2observations}).

This angular dependence limits the use of DM as a unique clock for the age of an FRB source. Specifically, interpreting a slope change with a spherical, single-density-profile model could assign it to the wrong physical interface, biasing estimates of the source age or surrounding density. Conversely, sustained monitoring that resolves curvature or a reversal in the DM evolution can provide more information than a single measurement of $d{\rm DM}/dt$. We note that other constraints could possibly
come from combining DM evolution with observables sensitive to different properties of the same plasma, including RM, temporal scattering, and the evolution of any persistent radio source, though we defer the discussion of these other phenomena in the AIC context to future work. 

For isotropically distributed viewing directions, the two polar cones occupy a fraction \(1-\cos\theta_{\rm cav}\simeq0.06\) for our adopted \(\theta_{\rm cav}=20^\circ\), while the remaining
\(\cos\theta_{\rm cav}\simeq0.94\) of sightlines are nonpolar. The bottom  of Figure~\ref{fig:Compared2observations} therefore represents the approximately 6\% polar class, whereas the top of Figure~\ref{fig:Compared2observations} illustrates the much more common nonpolar behavior; other nonpolar inclinations are qualitatively similar. Particularly when the upstream material is initially neutral, nonpolar sightlines encounter the additional wind-origin reservoirs and are therefore more likely to exhibit positive \(d{\rm DM}/dt\), and to retain it for longer, than polar sightlines. For an individual source, a polar-like DM history may indicate that the sightline passes through an evacuated cavity, but it could instead reflect a weaker or more diffuse wind structure. At the population level, finding substantially more Figure~\ref{fig:Compared2observations} bottom-like histories than the geometric expectation of $\sim$6\% would therefore favor either a larger effective polar opening angle or less prominent wind structure than adopted here. These geometric fractions are prior probabilities; angle-dependent detectability, including free--free absorption, could also bias the observed distribution of sightlines.

\subsection{The role of the fiducial parameters}
Here we adopted fiducial parameters of $t_w=5\times 10^3$~yr, $v_w=1000$~km~sec$^{-1}$, and $\dot{M}_w=10^{-6}$~M~$_{\odot}$~yr$^{-1}$. These are motivated by the wind of an interacting binary, such as the one following a double-white-dwarf merger, where the relevant outflow is the wind of the super-Chandrasekhar remnant during its luminous ($L\sim 3\times 10^4$~L$_\odot$) giant phase \citep[e.g.,][]{Schwab+16,Yu+19}. These choices also motivate where we place the external dense cloud. In particular, $t_w v_w\sim 5$~pc represents how far freely moving wind material could travel in the absence of confinement, but the interaction with the external material yields that the material swept up by the outer forward shock (see Eq.~\ref{eq:Rbw_appendix}), $R_{\rm wb}\sim 0.27$~pc ($R_{\rm wb}\sim 0.93$~pc), for $n_{\rm ext}=500$~cm$^{-3}$ ($n_{\rm ext}=1$~cm$^{-3}$). We therefore place the external material farther away ($R_{\rm out}$) from this value (see Table \ref{tab:modelparamters}). We tested a range of $t_w=10^{3-5}$~yr, and a range of $v_w=100-4000$~km~sec$^{-1}$ (revising $R_{\rm out}$ accordingly), and as expected, {{varying these parameters produces qualitatively similar DM histories, while changing the DM near flat value, the magnitude of $d{\rm DM}/dt$,
and the times of interface crossings and any sign reversals.
The intervals of rising and declining DM can therefore occur at different source ages for different parameter choices}} 
(as expected from the Equations (\ref{eq:Rbw_appendix})-(\ref{eq:tts_appendix}). We omit details of these tests here to avoid clutter.  

 Overall, the DM local maximum value of the toy model reflects the external environment. {{Specifically, the approximate plateau level of ${\rm DM}(t)$ in the ionized-upstream model and the local maximum reached after the rising phase in the initially neutral model (Figure~\ref{fig:Compared2observations}) reflect the electron column supplied by external origin material.
 }}
 The external material that is swept into the bubble's shell, i.e., DM$\approx n_{\rm ext}(R_{\rm out}-R_{\rm wb}) + n_{\rm ext}R_{\rm wb}/3$.  The \(n_{\rm ext}=500~{\rm cm^{-3}}\) is motivated by 
 \citet{Waxman+26} suggestion that the FRB~20121102A is confined by cold plasma on scales of \(0.03\)--\(0.3~{\rm pc}\). As expected, modifying the dense-cloud density mainly changes the DM normalization.

The ejecta properties associated with the collapse of a white-dwarf merger remnant are uncertain and depend sensitively on rotation, disk formation, and magnetic-field amplification. Neutrino-driven accretion-induced-collapse calculations produce weak explosions with \(M_{\rm ejc}\sim{\rm few}\times10^{-3}\,M_\odot\) and \(E_{\rm exp}\sim{\rm few}\times10^{49}\) erg, while outflows from a rotationally supported disk can increase the ejecta mass to \(\sim10^{-2}\)–few\(\times10^{-2}\,M_\odot\) \citep{Dessart+06,Metzger+09,Darbha+10,Yu+19}. At the energetic extreme, magnetically driven collapse calculations eject \(\sim0.1\,M_\odot\) with energies of order \(10^{51}\) erg or greater \citep[e.g.,][]{Dessart+07}.
We chose instead \((0.1\,M_\odot,10^{50}\,{\rm erg})\) as the upper limit, bracketing the ejecta mass/inertia at fixed velocity, since both of the upper and lower limit choices here  correspond to characteristic velocity of \(v_{\rm ch}=(2E_{\rm k}/M_{\rm ejc})^{1/2}\simeq0.033c\), so that the bulk ejecta remain subrelativistic. The $10^{51}$~erg yields a thicker band in all the plots and does not alter the qualitative results presented here. 



 As noted, we also adopt the ejecta power law with $n=10$ and $\delta=0$. We tested changing the power law in the range of $6-14$, and as expected the qualitative result remains, but the power laws of the variability change. 

\subsection{The role of triples}
The structure considered here is motivated not only by the AIC event itself, but also by the evolutionary pathways that lead to it. Using population models that include both isolated binaries and hierarchical triples, \citet{Shariat+26} found that triple dynamics enhances the formation of interacting WD systems via the Eccentric Kozai-Lidov mechanism \citep{Kozai,Lidov,Naoz16}, and opens merger pathways inaccessible to isolated binaries. Among \citet{Shariat+26} triple-origin FRB progenitor candidates, about 65\% undergo AIC through a single-degenerate channel involving a WD interacting with a nondegenerate companion, while about 35\% arise from double-WD mergers. The tertiary does not necessarily remain dynamically important, or even bound, at the time of collapse: its role is to excite the inner binary's eccentricity and thereby trigger tidal evolution, mass transfer, common-envelope evolution, or a merger. These processes naturally produce anisotropic mass loss and can shape the surrounding medium before the AIC. Triples therefore provide a natural evolutionary route to the central configuration assumed in our model: a newly formed neutron star embedded in an environment sculpted by prior binary interaction (illustrated in Figure \ref{fig:Cartoon}). Moreover, the long delay times from triple evolution allow the system to reside in an old stellar population, while the final episode of mass transfer or merger can still create a comparatively young, structured local environment shortly before collapse.

\subsection{Interpreting the toy model}
The model presented here is deliberately phenomenological. Its purpose is not to provide a hydrodynamically complete reconstruction of a particular AIC progenitor, but to isolate how a finite, anisotropic circumsource structure, upstream ionization, and viewing angle can shape the secular DM evolution. The predicted curves should therefore be interpreted as representative evolutionary pathways rather than precise source-age determinations. In particular, we regard the sequence of slope changes, the possibility of a sign reversal, and their viewing-angle dependence as more robust than the precise radii and times of individual transitions.

Several aspects of the binary outflow and AIC ejecta have been idealized. We represent the pre-AIC outflow by a smooth, axisymmetric $K(\theta)r^{-2}$ wind, whereas binary outflows may be spiral, clumpy, and time-dependent. Likewise, the AIC ejecta are represented by a single homologously expanding broken power law and a spheroidal angular structure, although realistic AIC outflows may contain multiple components and depend sensitively on rotation and magnetic fields \citep[e.g.,][]{Dessart+06,Batziou+25,Combi+25}. After the first environmental transition, we evolve each angular sector independently using a one-zone thin-shell prescription. This neglects lateral pressure forces, mixing between adjacent sectors, and reflected or transmitted waves at density interfaces; the reverse-shock coordinate should similarly be regarded as a mass-processing proxy rather than a resolved hydrodynamic shock. We also bracket the ionization physics using fixed prescriptions, rather than evolving photoionization, cooling, and recombination self-consistently.

The pre-AIC shocked wind is approximated as a uniform-density region whose density is fixed by mass conservation. Although \(R_{\rm wb}\) and \(P_{\rm b}\) follow the standard energy-driven bubble scalings and \(R_{\rm ts}\) is set by pressure balance, this closure does not reproduce the full radial density and temperature structure of an adiabatic wind bubble, particularly when \(R_{\rm ts}\) occupies a substantial fraction of the bubble. A more complete calculation would evolve the wind bubble, its termination shock, and the subsequent AIC-driven shell self-consistently. Consequently, the shocked-wind density and the precise transition times between circumsource zones should be treated as 
rough estimates only.
This approach affects the quantitative density normalization and transition times, but not the qualitative geometrical conclusion, i.e., that different sightlines traverse different material reservoirs, so the resulting DM evolution depends on viewing angle.

We have also placed the newly formed neutron star (the presumed FRB source) at the origin. A natal kick would displace it by
\begin{equation}
\Delta R_{\rm NS}\simeq
0.01\,{\rm pc}
\left(\frac{v_k}{100\,{\rm km\,s^{-1}}}\right)
\left(\frac{t}{100\,{\rm yr}}\right).
\end{equation}
Thus, a \(100~{\rm km~s^{-1}}\) kick produces a displacement of approximately \(0.01\) pc after \(100\) yr, while a \(400~{\rm km~s^{-1}}\) kick produces a displacement of approximately \(0.2\) pc after \(500\)~yr, for example. Such motion would break the simple correspondence between viewing inclination and a radial ray through the original center and could become important at late times, particularly in the compact-cloud model \citep[e.g.,][]{Ocker+26Kick}.

Finally, \(R_{\rm out}\) defines the outer boundary of the explicitly modeled external-material reservoir. After the forward shock passes this radius in the dense-cloud calculation, the dynamics continue into an \(n_{\rm bg}=1~{\rm cm^{-3}}\) background, whereas the plotted external-medium DM continues to track only material that was initially located inside \(R_{\rm out}\). Once the variable local contribution falls below \(\sim1~{\rm pc~cm^{-3}}\), the result becomes sensitive to the adopted extent and ionization state of this background; the dense-model evolution after $\sim10^3$~yr should therefore be viewed as illustrative. 

The rising DM branch is physically analogous to that discussed by \citet{Waxman+26}, in that shock processing converts previously neutral material into an observable electron column. The cause is nevertheless different: in their model the shock is powered by the relativistic persistent-radio-source plasma, whereas here it is driven by the AIC ejecta. 
Thus, we suggest that a rising DM is not unique to either engine; distinguishing these possibilities may require the DM evolution to be considered together with other observables such as possible persistent-source luminosity, RM, scattering, and other diagnostics of the surrounding plasma. 

The rapid dilution of the expanding ejecta favors early radio transparency. Using the thermal free–free absorption coefficient \citep[e.g.,][]{Zhao+21}, our adopted ejecta profile (\(n=10\), \(\delta=0\), \(q_{\rm core}=0.1\)), while the reverse shock remains outside the flat core, yields
\begin{eqnarray}
  \tau_{\nu,\rm unsh,ejc}^{\rm ff}&\simeq{}&  \left(\frac{1.3\,{\rm yr}}{t}\right)^5
\left(\frac{10^5\,{\rm km\,s^{-1}}}{v_{\rm ejc}(i)}\right)^5
\left(\frac{1\,{\rm GHz}}{\nu}\right)^2\nonumber\\
&\times&
\left(\frac{M_{\rm ejc}}{0.1\,M_\odot}\right)^2
f_\Omega(i)^2
\left(\frac{x_eY_e}{0.05}\right)^2  \\
&\times&
\left(\frac{T_e}{10^4\,{\rm K}}\right)^{-3/2}
\left(\frac{\bar g_{\rm ff}}{6}\right)
\mathcal Z_{\rm ff} \ , \nonumber
\end{eqnarray}
where \(f_\Omega(i)\equiv(4\pi/M_{\rm ejc})\,dM_{\rm ejc}/d\Omega\) describes the angular mass distribution, \(v_{\rm ejc}(i)\) is the outer-edge ejecta velocity, \(\bar g_{\rm ff}\) is the Gaunt factor, and \(\mathcal Z_{\rm ff}\equiv\sum_a Z_a^2n_a/n_e\) accounts for ionic charges. Both \(t\) and \(\nu\) are evaluated in the source frame. The \(t^{-5}\) scaling follows from the squared density dilution, \(t^{-6}\), multiplied by the expanding path length, \(t\), at fixed thermal and ionization properties. For \(T_e=10^4\) K, \(\bar g_{\rm ff}=6\), \(\mathcal Z_{\rm ff}=1\), and our adopted \(x_eY_e=0.05\), the unshocked ejecta become transparent at \(1\) GHz after approximately \(0.5\) yr and \(1.2\)–\(1.3\) yr for the weak and energetic fiducial models, respectively. These estimates place ejecta transparency well before the decade-to-century DM reversals highlighted here, supporting the possibility of observing these later signatures. The detailed visibility window depends on the thermal and ionization histories, ionic composition, and density substructure, together with absorption and pulse scattering in the surrounding zones.

\subsection{Comparison with observed repeaters}\label{sec:Compare2Observations}
In Figure \ref{fig:Compared2observations} we overplot a few representative observations of $d{\rm DM}/dt$.  Values for FRB 20190117A, FRB 20190907A, FRB 20210323C are estimated from \citet{Cook+26}, and FRB 121102, FRB 20240619D, FRB 20220912A and  FRB 20190520B estimated from \citet{Hessels2019,Ould-Boukattine+26,Abbott+26FRB} and \citet{Niu+26}, respectively; see the Figure caption for details. 

FRB~121102 is a good test case, because its DM rose for several years before declining \citep[e.g.,][]{Spitler2016,Scholz2016,Hessels2019,Oostrum2020,Majid2020,Platts2021,Snelders2023,Wang2025,Zhang2025}. In our framework, such a history requires the forward shock to have crossed a dense, initially neutral medium. 
The rising branch in our dense-cloud model shares its underlying
microphysics with that of \citet{Waxman+26}: in both models, shock
processing converts an initially neutral, finite reservoir into an
observable electron column. The similarity between their inferred age
of approximately \(60\)~yr and our rising phase at
\(t\sim10\)--\(10^2\)~yr is therefore an order-of-magnitude consistency
rather than a discriminating prediction. The physical drivers and
geometries are nevertheless different. In their model, the shock is
continuously driven by the relativistic plasma responsible for the
persistent radio source, whereas here it is driven by the impulsively
launched AIC ejecta. Their finite-shell model naturally produces one
principal rise--breakout--decline sequence, while our structured
circumsource environment can introduce additional slope changes and
short-lived excursions as successive interfaces are crossed, with
transition times that depend on viewing angle. {{
The connection to \citet{Waxman+26} concerns the shock ionization of initially neutral material, rather than a necessary association with detectable persistent radio emission. In our framework, the AIC ejecta drive the shock, so the dense-cloud DM evolution does not require a detectable persistent radio source. Such emission is not excluded, but predicting its luminosity would require an additional model for the energetic particles, magnetic fields, and radiative transfer, beyond the scope of this work. }}


\begin{figure*}
    \centering
    \includegraphics[width=0.99\linewidth]{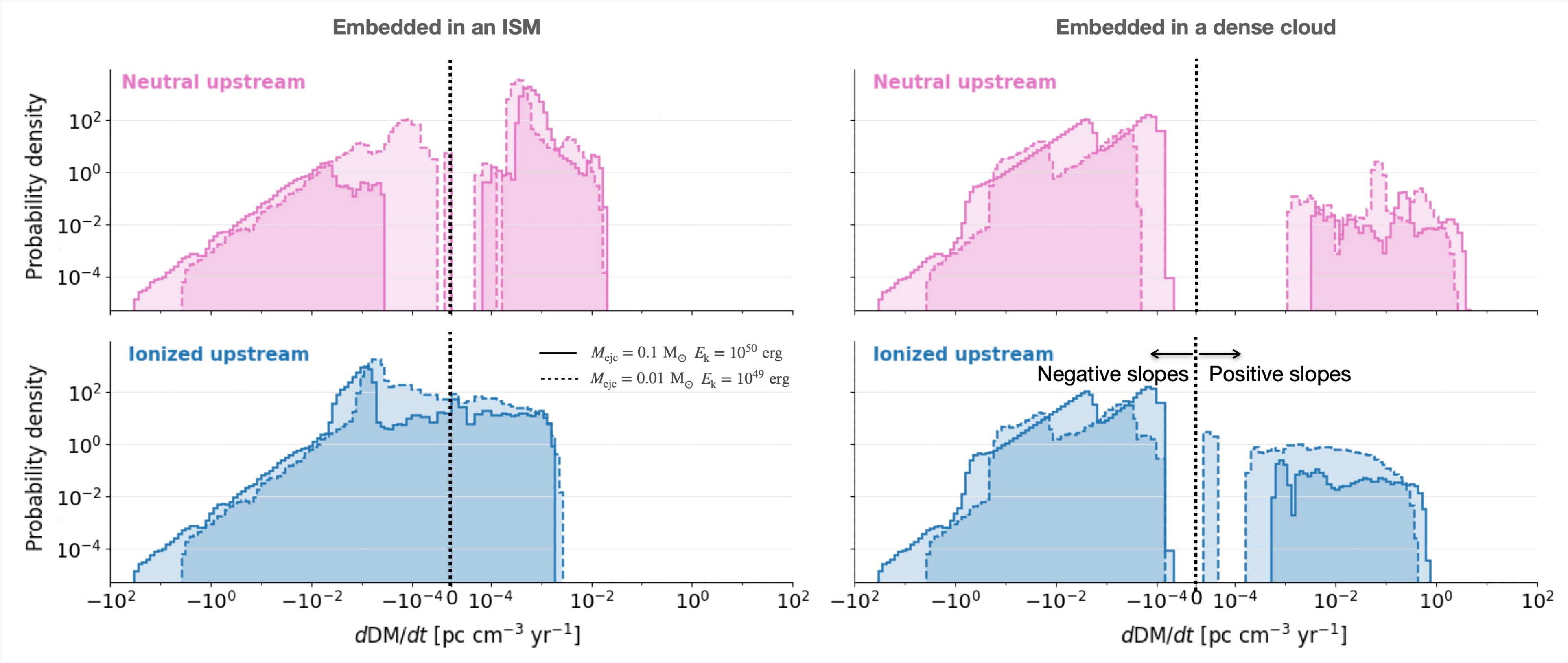}
    \caption{Probability distributions of the signed secular host-frame DM slope, $d{\rm DM}/dt$, for mock populations of repeating FRB sources. The left column shows the diffuse external-medium model, $n_{\rm ext}=1~{\rm cm^{-3}}$ and $R_{\rm out}=10~{\rm pc}$, whereas the right column shows the finite dense-cloud model, $n_{\rm ext}=500~{\rm cm^{-3}}$ and $R_{\rm out}=0.3~{\rm pc}$.
The external material is initially neutral in the upper row (pink) and fully ionized in the lower row (blue); material processed by the AIC forward shock is assumed to be ionized in both cases.
For each panel, we draw $10^4$ mock sources with ages distributed uniformly in linear time between $1$ and $10^3$~yr. Viewing directions are drawn isotropically. We approximate the angular dependence using representative non-polar and polar sightlines, $i=90^\circ$ and $i=0^\circ$, weighted by $\cos\theta_{\rm cav}=0.94$ and $1-\cos\theta_{\rm cav}=0.06$, respectively, for a polar-cavity half-opening angle $\theta_{\rm cav}=20^\circ$.
For each source, we generate 20 DM measurements with exponentially distributed inter-detection intervals having a mean of $135$~days, corresponding to a mean monitoring baseline of approximately $7$~yr, and obtain one value of DM slope $S$ from a linear fit to ${\rm DM}(t)$. Solid curves correspond to $(M_{\rm ej},E_K)=(0.1\,M_\odot,10^{50}\,{\rm erg})$, and dashed curves to $(0.01\,M_\odot,10^{49}\,{\rm erg})$. {{The scenario in which the FRB is embedded in an ISM and has a neutral upstream may yield a large abundance of positive slopes (specifically 96\% and 94\% for the solid and dashed lines, respectively), but they are limited to low values $\lsim 0.015$~pc~cm$^{-3}$~yr$^{-1}$. The ISM for an ionized upstream is far more limited in its possible slopes $\lsim 0.002$~pc~cm$^{-3}$~yr$^{-1}$, and far less frequently ($\sim 2\%$ and $3\%$) for the solid and dashed lines, respectively.  On the other hand, in the scenario where the FRB is embedded in a dense cloud, the slope may reach higher values of $\sim 1$~pc~cm$^{-3}$~yr$^{-1}$ and $4$~pc~cm$^{-3}$~yr$^{-1}$ for the neutral and ionized upstream limits, respectively. The frequency of the positive slopes varies from $5\%$ ($18\%$) to $2\%$ ($5\%$) for the solid (dashed) lines in the neutral and ionized upstream limits, respectively. }}
The same mock ages, orientations, and observing schedules are used for all physical models. Each curve is normalized independently so that its area
over the complete signed-slope range is unity, where the density is defined per unit linear $S$ despite the symmetric-logarithmic horizontal scale.
The vertical axis is logarithmic, and the dotted line marks $S=0$, separating declining from increasing DM histories. The translucent regions show the area beneath each model distribution; darker overlap between the solid and dashed distributions does not represent an uncertainty interval. Measurement errors and observational selection are not included. }
    \label{fig:SlopesHist}
\end{figure*}

Notably, \citet{Metzger26} suggested that DM variations are driven by changes in photoionization rather than by dynamics. The two pictures make different predictions: photoionization ties the DM to the engine's time-averaged ionizing luminosity, whereas in our model the DM responds to the shock kinematics and to the radial structure of the
environment, independent of the burst activity. The dynamical rise also requires the shocked gas to remain ionized. For hydrogen-rich upstream gas and $v_{\rm fs}\gtrsim10^3\ \mathrm{km\,s^{-1}}$ the post-shock temperature is $\gtrsim10^7$\,K and the cooling time is $\gtrsim10^3$\,yr for $n_{\rm ext}\sim500\ \mathrm{cm^{-3}}$, longer than the rising phase; for
slower shocks or metal-rich upstream gas (e.g., merger debris), cooling and
recombination can be rapid, as emphasized by \citet{Metzger26}, and the
rise we find should be regarded as an upper limit.

A potentially observable distinction between these interpretations is
their relation to FRB activity. In our dynamical framework,
\({\rm DM}(t)\) is determined by shock propagation through the
circumsource density and ionization structure and is therefore
independent of the contemporaneous burst rate. By contrast, the
photoionization model of \citet{Metzger26} connects changes in the
ionized column to the time-dependent high-energy output of the central
engine and predicts that enhanced activity should be followed by a
temporally smoothed DM increase. A reproducible association between
activity episodes and subsequent DM increases would therefore favor
photoionization, whereas 
a reversal in the
DM slope without a corresponding change in burst activity would favor
a dynamical origin.

Notably, Figure \ref{fig:Compared2observations} shows that, without fitting any individual source, our framework spans both signs and the approximate magnitudes of the secular DM-variation rates measured for repeating FRBs. Positive, negative, and sign-changing values of $d{\rm DM}/dt$ can all arise within the same physical model as the forward shock traverses successive circumsource regions and the balance between ionization, mass sweeping, and geometric dilution changes. The observed diversity of $d{\rm DM}/dt$ therefore does not necessarily imply a corresponding diversity of central engines; it may instead reflect different source ages, external densities, ionization states, and viewing angles. 

{{The duration of a DM decline provides information beyond its instantaneous slope. In our fiducial diffuse-medium model, the steepest negative slopes occur during the first few years after AIC, implying that the FRB activity may commence shortly after the AIC formed. The unshocked ejecta can already be transparent at GHz frequencies during part of this interval, but the predicted decline rapidly flattens. In the dense-cloud model, dilution of the swept-up cloud material can sustain a decline at later ages, although the fiducial cloud does not reproduce the steepest observed rates at those ages. A sustained, rapid decline thus favors a substantial, evolving local electron column in that source; a negative slope alone does not require a dense external medium.
}}

\subsection{The distribution of secular DM slopes}
As more repeating FRBs are discovered \citep[e.g.][]{Cook+26}, it is interesting to consider on a population level the statistics of DM variations expected in our model.
Figure~\ref{fig:SlopesHist} shows the distribution of secular DM slopes for a mock population of sources with different ages and viewing angles. We generate $10^4$ mock sources with ages sampled uniformly between $1$ and $10^3$~yr, approximating isotropically distributed viewing directions by polar and nonpolar solutions with their respective geometric weights of approximately
$6\%$ and $94\%$. For each source, we fit a straight line to 20 synthetic DM measurements obtained at irregular intervals over an average baseline of approximately $7$~yr. 
Specifically, for each source, the first synthetic DM measurement is made at its randomly assigned age. We generate the subsequent 19 measurement times by drawing independent waiting times from an exponential distribution with a mean of $135$~days, corresponding to a constant-rate Poisson process for detected bursts \citep[although we note that most very active repeaters show clear burst clustering, e.g.][]{lac+22,Abbott+26FRB}. This mean interval is chosen so that the 20 measurements span approximately $7$~yr on average, $\langle T_{\rm obs}\rangle=19\times135~{\rm days}\simeq7$~yr,
although the monitoring duration varies between sources. We then fit a straight line to these measurements to obtain one secular DM slope per source. This prescription provides a simple model of irregular
bursting;
 the resulting distributions therefore describe finite-baseline, estimated slopes rather than instantaneous derivatives.

As seen in Figure~\ref{fig:SlopesHist}, for both ejecta models, the large-positive-slope value tail occurs only in the dense-cloud scenarios explored here, whereas the diffuse-medium models produce much weaker increases. Large positive slopes thus favor a dense surrounding reservoir within our framework. Negative slopes, however, occur in all four combinations of environment and upstream ionization: a declining DM alone does not identify the surrounding medium.
Another notable feature is the depletion of slopes near zero in the initially neutral cases, particularly for the dense cloud, separating the rising and declining populations.

{{
Motivated by FRB~20121102A, we ask how often a single monitoring
campaign captures a DM rise followed by a decline, irrespective of amplitude. We count a turnover when the sampled DM increases over two consecutive intervals and decreases over the next two. Across both ejecta masses and both upstream ionization states, $\sim 0.5\%$--$0.7\%$ of the dense-cloud realizations meet this criterion. This fraction has a simple interpretation: for starting ages distributed uniformly from 1 to $10^3$~yr, a seven-year campaign has only a $\sim 7/(1000-1)\simeq 0.7\%$ chance of straddling a given DM maximum away from the age boundaries. In the diffuse medium with initially ionized gas, the rise--decline fractions are $1\%$ and $0.5\%$ for $M_{\rm ejc}=0.1\,M_\odot$ and $0.01\,M_\odot$, respectively; a near-zero fraction occurred in the $10^4$ realizations per mass with initially neutral diffuse gas. If we also count the opposite sequence, a decline followed by a rise, the dense-cloud fraction increases to approximately $1\%$ in all four cases. The $\sim 1\%$ here is encouraging given the nature of the toy model and the $\sim 10\%$ of observed sources with DM evolution turnovers. However, we do not expect many more such turnovers to be observed. 
}}

Observed distributions of secular DM slopes could therefore help distinguish between environmental and ionization scenarios while constraining the relative contributions of polar and nonpolar sightlines.
The informative features include not only the fraction of positive slopes, but also their magnitudes, the negative tail, and the prevalence of nearly stationary DM histories. These comparisons should account for the source-age distribution, monitoring baseline, measurement uncertainties, and angle-dependent detectability, which can modify the relative weights and blur the near-zero depletion. Population-level monitoring would thus complement the information obtained from the long-term evolution of individual repeaters.

Additionally, the \citet{Ocker+26Kick} model implies as many negative as positive secular DM slopes, while relative numbers in the toy model presented here depend on the FRB source activity as a function of age.  For example, if FRB activity occurs mostly shortly after the AIC is launched, we expect mostly negative slopes, while uniform activity yields the distributions shown in Figure \ref{fig:SlopesHist}. Therefore, we suggest the slope distribution offers a unique discriminator between models and ages.

\subsection{Summary}
We have shown that an FRB source formed through AIC in an interacting binary may be embedded in a finite, anisotropic circumsource environment sculpted by the progenitor's prior mass loss. Within our toy model, the interfaces between the resulting circumsource zones produce breaks and curvature in \({\rm DM}(t)\), while the upstream ionization state determines whether the total DM declines, rises, or reverses direction.  Because different sightlines encounter different material reservoirs at different times, an instantaneous measurement of \(d{\rm DM}/dt\) cannot by itself provide a unique source age. The same progenitor framework can therefore generate a broad range of DM histories through differences in environmental structure, ionization state, and viewing geometry. 

\section*{Acknowledgments}
S.N. and V.M.K. thank the organizers of the 8th edition of the workshop ``Unsolved Problems in Astrophysics and Cosmology,'' held from 6 to 10 July, 2026, in Klosterneuburg, Austria, which helped inspire the project. 
Generative Artificial Intelligence, ChatGPT, Claude, and Grammarly were used to help with brainstorming, editing, and formatting; the substantive ideas, analysis, and predictions presented are the team’s and were not generated by any AI tool.
S.N. thanks the support of NASA grant No. 80NSSC24K0773 (ATP-23-ATP23-0149) and Howard and Astrid Preston for their generous support.  V.M.K. holds the Lorne Trottier Chair in Astrophysics \& Cosmology, a Distinguished James McGill Professorship, and receives support from an NSERC Discovery grant (RGPIN 228738-13).



\appendix

\section{Dynamical prescription}\label{app:dynamics}

The DM components of Section \ref{sec:zones} require, at each time $t$ and viewing angle $\theta$, the radii of the AIC reverse shock, contact discontinuity, and forward shock ($R_{\rm rs}, R_{\rm cd}, R_{\rm fs}$), together with the pre-AIC interfaces ($R_{\rm ts}, R_{\rm cw}, R_{\rm wb}$) that they encounter. In this Appendix we describe how these radii are computed, first for the pre-AIC wind bubble (\S A.1) and then for the early interaction of the AIC ejecta with the free wind (\S A.2); the subsequent thin-shell evolution is given in Appendix \ref{app:thin_shell}.

We evolve each polar angle as an independent radial sector.  Before the first environmental interface is encountered, the non-polar regimes follow the self-similar solution of \citet{Chevalier82}.  Thereafter, we integrate a one-zone thin-shell model that conserves the swept mass and radial momentum and follows the thermal energy generated by the shocks.  This construction retains the leading dependence on viewing angle but neglects lateral pressure
gradients and exchange of mass or energy between neighboring sectors. 

\subsection{Pre-AIC wind bubble}\label{app:pre_aic_bubble}

For the angle-independent wind speed adopted here, the mechanical luminosity
of the pre-AIC wind is $L_{\rm w}=\dot{M} v_{\rm w}^{2}/2$. 
In the energy-conserving wind-bubble solution \citep{Castor+75,Weaver+77}, the outer bubble shock and the approximately uniform interior pressure at the time of AIC are 
\begin{align}
    R_{\rm wb}(t_{\rm w})
      &= \left(\frac{125}{154\pi}\right)^{1/5}
         \left(\frac{L_{\rm w}}{\rho_{\rm ext}}\right)^{1/5}
         t_{\rm w}^{3/5},
         \label{eq:Rbw_appendix}\\
    P_{\rm b}(t_{\rm w})
      &= \frac{7}{(3850\pi)^{2/5}}
         L_{\rm w}^{2/5}\rho_{\rm ext}^{3/5}t_{\rm w}^{-4/5}.
         \label{eq:Pb_appendix}
\end{align}
Here $\rho_{\rm ext}=m_{\rm p}n_{\rm ext}$ is the density into which the pre-AIC bubble expanded.  We set the pre-AIC contact discontinuity to $ R_{\rm cw}=\xi R_{\rm wb}$ with $\xi=0.86$.  The value of $\xi$ is a geometric closure for the toy model adopted from  \citet{Castor+75Wind} and \citet{Weaver+77}. 

Outside the polar cavity, the freely expanding wind has
$\rho_{\rm w}=K(\theta)r^{-2}$, with
$K(\theta)=[d\dot M/d\Omega(\theta)]/v_{\rm w}$ (see Eq.~\ref{eq:Ktheta}).  Balancing its ram pressure
against $P_{\rm b}$ gives the wind-termination radius
\begin{equation}
    R_{\rm ts}(\theta)
      =\left[\frac{K(\theta)v_{\rm w}^{2}}{P_{\rm b}}\right]^{1/2}.
    \label{eq:Rts_appendix}
\end{equation}
There is no wind termination shock inside the imposed polar cavity.

For completeness, the mass of pre-AIC shocked-wind material per unit solid
angle is
\begin{equation}
    \frac{dM_{\rm sw}}{d\Omega}
      =\frac{d\dot M}{d\Omega}t_{\rm w}
       -K(\theta)\left[R_{\rm ts}(\theta)-R_{\rm w,in}\right],
    \label{eq:Msw_appendix}
\end{equation}
and our uniform-density closure gives
\begin{equation}
    \rho_{\rm sw}(\theta)
      =\frac{3\,dM_{\rm sw}/d\Omega}
             {R_{\rm cw}^{3}-R_{\rm ts}^{3}(\theta)}.
    \label{eq:rhosw_appendix}
\end{equation}
Likewise, placing all external-medium mass initially enclosed by
$R_{\rm wb}$ into the layer $R_{\rm cw}<r<R_{\rm wb}$ gives
\begin{align}
    \frac{dM_{\rm shell}}{d\Omega}
      &=\frac{\rho_{\rm ext}R_{\rm wb}^{3}}{3},
      \label{eq:Msh_appendix}\\
    \rho_{\rm shell}
      &=\frac{\rho_{\rm ext}}
              {1-\left(R_{\rm cw}/R_{\rm wb}\right)^3}.
      \label{eq:rhosh_appendix}
\end{align}
Equations~\eqref{eq:rhosw_appendix} and \eqref{eq:rhosh_appendix} conserve
the mass in each sector, but do not reproduce the internal radial structure of
a resolved wind bubble.

The dense-cloud calculation also follows the dynamics after the shock leaves
the explicitly modeled cloud.  Its ambient density is
\begin{equation}
  \rho_{\rm amb}(r)=
  \begin{cases}
    \rho_{\rm ext}, & r<R_{\rm out},\\
    \rho_{\rm bg},  & r\geq R_{\rm out},
  \end{cases}
  \label{eq:ambient_density_appendix}
\end{equation}
where the dense cloud scenario of  $n_{\rm ext}=500\,\mathrm{cm^{-3}}$,
$R_{\rm out}=0.3\,\mathrm{pc}$, and
$n_{\rm bg}=1\,\mathrm{cm^{-3}}$.  In the diffuse calculation,
$n_{\rm ext}=n_{\rm bg}=1\,\mathrm{cm^{-3}}$, so this boundary does not
produce a density jump.

\subsection{Early interaction with the free wind}\label{app:early_wind}

Before the AIC forward shock reaches $R_{\rm ts}(\theta)$, a nonpolar sector
follows the $s=2$ ejecta-dominated similarity solution.  Writing
$\rho_{\rm w}=K(\theta)r^{-2}$,
\begin{align}
    R_{\rm cd}(t,\theta)&=C_2(\theta)t^{m_2},
        \label{eq:Rcd_early}\\
    C_2(\theta)&=
      \left[\frac{A_2g_n(\theta)}{K(\theta)}\right]^{1/(n-2)},
        \label{eq:C2_early}\\
    m_2&=\frac{n-3}{n-2} \ ,
        \label{eq:m2_early} 
\end{align}
where $ R_{\rm fs}=\beta_{\rm fs}^{(2)}R_{\rm cd}$,  $
    R_{\rm rs}=\beta_{\rm rs}^{(2)}R_{\rm cd}$ and
\begin{equation}\label{eq:ms}
    m_s = \frac{n-3}{n-s} \ ,
\end{equation}
thus, $m_2=m_{s=2}$.
The constants $A_2$, $\beta_{\rm fs}^{(2)}$, and
$\beta_{\rm rs}^{(2)}$ are taken from Table~1 of
\citet{Chevalier82}; for $n=10$ and $\gamma=5/3$ we use
$A_2=0.067$, $\beta_{\rm fs}^{(2)}=1.239$, and
$\beta_{\rm rs}^{(2)}=0.984$.  The forward shock reaches the wind
termination shock at
\begin{equation}
    t_{\rm ts}(\theta)=
      \left[\frac{R_{\rm ts}(\theta)}
      {\beta_{\rm fs}^{(2)}C_2(\theta)}\right]^{1/m_2}.
    \label{eq:tts_appendix}
\end{equation}

Inside the polar cone, there is no corresponding wind-interaction stage.  The ejecta expand freely until their outer edge reaches the cavity wall, $  t_{\rm hit}(\theta)={R_{\rm cav}}/{v_{\rm ejc}(\theta)}$. 
Throughout the polar cone, for the fiducial choice used here,
$R_{\rm cav}=R_{\rm wb}$.

\section{Post-transition thin-shell evolution}\label{app:thin_shell}

The self-similar solution described in Appendix~A.2 applies while the outer ejecta interact with the scale-free, freely expanding wind. This approximation fails when 
the forward shock reaches the wind termination shock, or once the polar ejecta strike the cavity wall, because the shock subsequently encounters finite layers with abrupt changes in density and density slope. We therefore evolve the post-transition flow in each angular sector using a one-zone thin-shell model. The shell gains mass and radial momentum from the homologously expanding ejecta at its inner boundary and from the approximately stationary pre-AIC material at its outer boundary. Kinetic energy dissipated at these boundaries is deposited as thermal energy; the resulting pressure accelerates the shell, while \(P\,dV\) work accounts for adiabatic expansion. This prescription captures the changing inertia and pressure forces as the shell crosses \(R_{\rm cw}\), \(R_{\rm wb}\), and \(R_{\rm out}\), but does not resolve the internal structure of the shocked layers, lateral flows between angular sectors, or reflected and transmitted waves at the interfaces.

At \(t_{\rm ts}\) in a nonpolar sector, or at \(t_{\rm hit}\) in a polar sector, we initialize the thin-shell integration. Outside the polar cavity, the forward-shock and reverse-shock-proxy radii are related to the contact discontinuity by
\(R_{\rm fs}=\beta_{\rm fs}^{(2)}R_{\rm cd}\) and
\(R_{\rm rs,p}=\beta_{\rm rs}^{(2)}R_{\rm cd}\).
These ratios are inherited from the early \(s=2\) Chevalier solution and retained as geometric lead factors throughout the post-transition integration; they do not imply that the later evolution remains self-similar. Inside the polar cavity we set both factors to unity, corresponding to a geometrically thin shell. The quantity \(R_{\rm rs,p}\) should therefore be interpreted as the mass-processing boundary used by the one-zone model rather than as a resolved hydrodynamic reverse-shock trajectory.



We denote the shocked-ejecta and swept pre-AIC masses per unit solid angle by $\mu_{\rm ejc}$ and $\mu_{\rm pre}$, respectively.  The freely expanding ejecta entering the reverse-shock proxy have velocity $u_{\rm ejc}={R_{\rm rs,p}}/{t}$. We take the pre-AIC material to be stationary, $u_{\rm pre}=0$.  The mass loading rates are
\begin{align}
    \dot\mu_{\rm ejc}
       &=\rho_{\rm ejc}(R_{\rm rs,p},\theta,t)
         R_{\rm rs,p}^{2}\left(u_{\rm ejc}-V_{\rm rs,p}\right),
         \label{eq:mdot_ejecta_shell}\\
    \dot\mu_{\rm pre}
       &=\rho_{\rm pre}(R_{\rm fs},\theta)
         R_{\rm fs}^{2}V_{\rm fs} \ ,
         \label{eq:mdot_pre_shell}
\end{align}
where \(V_{\rm fs}=\beta_{\rm fs}^{(2)}V_{\rm cd}\) and \(V_{\rm rs,p}=\beta_{\rm rs}^{(2)}V_{\rm cd}\) outside the polar cone. Both factors are unity in the polar calculation.
Equation~\eqref{eq:mdot_ejecta_shell} is used while
$u_{\rm ejc}>V_{\rm rs,p}$; if the proxy no longer propagates back through the homologous ejecta in velocity coordinates, we set $\dot\mu_{\rm ejc}=0$.
We also stop ejecta loading once the full ejecta mass in that sector has been processed.

We adopt adiabatic shocks and solve the following equations: 
\begin{align}
    \frac{dR_{\rm cd}}{dt} &= V_{\rm cd},
        \label{eq:thin_shell_radius}\\
    \frac{d\mu_{\rm ejc}}{dt} &= \dot\mu_{\rm ejc},\\ 
    \frac{d\mu_{\rm pre}}{dt} &= \dot\mu_{\rm pre},
        \label{eq:thin_shell_masses}\\
    (\mu_{\rm ejc}+\mu_{\rm pre})\frac{dV_{\rm cd}}{dt}
      &=\dot\mu_{\rm ejc}(u_{\rm ejc}-V_{\rm cd})
       +\dot\mu_{\rm pre}(u_{\rm pre}-V_{\rm cd})
       +P R^{2},
        \label{eq:thin_shell_momentum}\\
    \frac{dU}{dt}
      &=\frac{1}{2}\dot\mu_{\rm ejc}(u_{\rm ejc}-V_{\rm cd})^2
       +\frac{1}{2}\dot\mu_{\rm pre}(u_{\rm pre}-V_{\rm cd})^2
       -P R_{\rm cd}^{2}V_{\rm cd},
        \label{eq:thin_shell_energy}\\
    P&=(\gamma-1)\frac{U}{R_{\rm cd}^{3}/3},\qquad \gamma=\frac{5}{3}.
        \label{eq:thin_shell_pressure}
\end{align}
All masses and energies in these equations are per unit solid angle, so the sector volume is $R_{\rm cd}^{3}/3$ and the pressure force is $P R_{\rm cd}^2$.  The first two terms in Equation~\eqref{eq:thin_shell_energy} thermalize the relative kinetic energy of newly processed ejecta and external material; the last term is the adiabatic $P\,dV$ loss.  The pre-existing pressure $P_{\rm b}$ is used to set $R_{\rm ts}$ but is not added as a separate upstream pressure term in the
post-AIC shell equation.

\subsection{Initial conditions}\label{app:shell_initial_conditions}

For a nonpolar sector, the integration begins at $t_*=t_{\rm ts}$. 
The initial conditions inherited from the similarity solution are $R_*={R_{\rm ts}}/{\beta_{\rm fs}^{(2)}}$, $V_*=m_2{R_*}/{t_*}$ and $q_*={\beta_{\rm rs}^{(2)}R_*}/({v_{\rm ejc}t_*}$), as well as:
\begin{eqnarray}
     \mu_{{\rm ejc},*}
      &=& \frac{dM_{\rm ejc}}{d\Omega}
        \frac{I_2(q_*,1)}{I_2(0,1)}, \\
      \mu_{{\rm pre},*}
      &=&K(\theta)\left(R_{\rm ts}-R_{\rm w,in}\right),
      \label{eq:nonpolar_initial_masses}\\
    E_{{\rm proc},*}
      &=&\frac{1}{2}\frac{dM_{\rm ejc}}{d\Omega}v_{\rm ejc}^{2}
        \frac{I_4(q_*,1)}{I_2(0,1)}, \\
      U_*&=&E_{{\rm proc},*}
        -\frac{1}{2}(\mu_{{\rm ejc},*}+\mu_{{\rm pre},*})V_*^2 \ .
      \label{eq:nonpolar_initial_energy}
\end{eqnarray}

For a polar sector, the shell is initialized when the freely expanding ejecta
strike the cavity wall: $t_*=t_{\rm hit}$ and $R_*=R_{\rm cav}$.  Let
$\rho_{{\rm ejc},*}$ and $\rho_{{\rm amb},*}$ denote the densities immediately
inside and outside that interface.  Equating the two cold-fluid ram pressures
gives the initial contact speed
\begin{equation}
    V_*=
    \frac{\sqrt{\rho_{{\rm ejc},*}}\,v_{\rm ejc}
          +\sqrt{\rho_{{\rm amb},*}}\,u_{\rm pre}}
         {\sqrt{\rho_{{\rm ejc},*}}+
          \sqrt{\rho_{{\rm amb},*}}}
    =\frac{\sqrt{\rho_{{\rm ejc},*}}\,v_{\rm ejc}}
           {\sqrt{\rho_{{\rm ejc},*}}+
            \sqrt{\rho_{{\rm amb},*}}}.
    \label{eq:polar_initial_velocity}
\end{equation}
Because no shell exists before impact, we seed the integration over $\Delta t=f_{\rm seed}t_*$, with $f_{\rm seed}=10^{-4}$, $ \dot\mu_{{\rm ejc},*}
      =\rho_{{\rm ejc},*}R_*^2(v_{\rm ejc}-V_*)$,
      $\dot\mu_{{\rm pre},*}
      =\rho_{{\rm amb},*}R_*^2V_*$,      
      $\mu_{{\rm ejc},*}=\dot\mu_{{\rm ejc},*}\Delta t$, $\mu_{{\rm pre},*}=\dot\mu_{{\rm pre},*}\Delta t$ and 
      \begin{equation}
          U_*=\frac{\Delta t}{2}
      \left[\dot\mu_{{\rm ejc},*}(v_{\rm ejc}-V_*)^2
      +\dot\mu_{{\rm pre},*}V_*^2\right]
      \end{equation}
The integration is divided into segments whose endpoints coincide with
$R_{\rm cw}$, $R_{\rm wb}$, and, in the dense model, $R_{\rm out}$.  The
state variables $(R,V,\mu_{\rm ejc},\mu_{\rm pre},U)$ remain continuous at each
boundary; the subsequent response to a density jump follows from
Equations~\eqref{eq:mdot_pre_shell}--\eqref{eq:thin_shell_energy}.  We do not
apply an instantaneous ram-pressure velocity reset at these interfaces.

\subsection{Reverse-shock proxy and core entry}\label{app:reverse_shock}

Before polar impact, all ejecta in that sector are unshocked.  At other times,
the reverse-shock coordinate used by the DM calculation is
\begin{equation}
 q_{\rm rs}(t,\theta)=
 \begin{cases}
  1, & 0\leq\theta\leq\theta_{\rm cav}\ \text{and}\ t\leq t_{\rm hit},\\
  \min\!\left[1,\dfrac{R_{\rm rs,p}(t,\theta)}
                         {v_{\rm ejc}(\theta)t}\right], & \text{otherwise}.
 \end{cases}
 \label{eq:qrs_appendix}
\end{equation}
The corresponding processed ejecta mass is
\begin{equation}
    \frac{dM_{\rm ej,sh}}{d\Omega}
      =\frac{dM_{\rm ejc}}{d\Omega}
       \frac{I_2(q_{\rm rs},1)}{I_2(0,1)}.
    \label{eq:shocked_ejecta_mass_appendix}
\end{equation}
We record the core-entry time from
$q_{\rm rs}(t_{\rm core})=q_{\rm core}$.  In the implementation used here,
the proxy relation mentioned above continues after core entry; we impose no finite time at which the reverse shock reaches the ejecta center.  Accordingly, $t_{\rm core}$ is a diagnostic landmark, whereas a separate $t_{\rm rs,center}$ is not part of the present calculation.  A more realistic post-core reverse-shock trajectory would require an additional closure or a hydrodynamic calculation.

\bibliography{Refs}

\end{document}